\documentclass[usegraphicx,usenatbib,useAMS]{mn2e}\usepackage{times}
  \let\url\relax

\newcommand{\be}{\begin{equation}}
\newcommand{\ba}{\begin{eqnarray}}
\newcommand{\ee}{\end{equation}}
\newcommand{\ea}{\end{eqnarray}}  

\def\lesssim{\mathrel{\hbox{\rlap{\hbox{\lower4pt\hbox{$\sim$}}}\hbox{$<$}}}}
\def\gtrsim{\mathrel{\hbox{\rlap{\hbox{\lower4pt\hbox{$\sim$}}}\hbox{$>$}}}}

\def\gtsima{$\; \buildrel > \over \sim \;$}
\def\ltsima{$\; \buildrel < \over \sim \;$}
\def\gsim{\lower.5ex\hbox{\gtsima}}
\def\lsim{\lower.5ex\hbox{\ltsima}}
\def\simgt{\lower.5ex\hbox{\gtsima}}
\def\simlt{\lower.5ex\hbox{\ltsima}}
\def\simpr{\lower.5ex\hbox{\prosima}}
\def\la{\lsim}
\def\ga{\gsim}

\newcommand{\hMpc}{h^{-1}\,{\rm Mpc}} \newcommand{\hMsun}{h^{-1}\,M_\odot}
 
\newcommand{\hmsun}{{\,\rm h^{-1}M}_\odot}
\newcommand{\hmpc}{{\,\rm h^{-1}Mpc}}

\def\simless{\mathbin{\lower 3pt\hbox
   {$\rlap{\raise 5pt\hbox{$\char'074$}}\mathchar''7218$}}}   
\def\simgreat{\mathbin{\lower 3pt\hbox
   {$\rlap{\raise 5pt\hbox{$\char'076$}}\mathchar''7218$}}}   

\begin{document}

\title  [Reionization of the Local Group] 
{Reionization of the Local Group of Galaxies} 
 
\author[I. T. Iliev, et al.]{Ilian~T.~Iliev$^{1,2}$\thanks{e-mail: 
I.T.Iliev@sussex.ac.uk}, Ben Moore$^{2}$, Stefan Gottl\"ober$^{3}$, 
Gustavo Yepes$^{4}$, Yehuda Hoffman$^{5}$ \newauthor and 
Garrelt Mellema$^{6}$
\\
$^1$ Astronomy Centre, Department of Physics \& Astronomy, Pevensey II 
Building, University of Sussex, Falmer, Brighton BN1 9QH, United Kingdom\\    
$^2$ Universit\"at Z\"urich, Institut f\"ur Theoretische Physik,
         Winterthurerstrasse 190, CH-8057 Z\"urich, Switzerland\\
$^3$ Astrophysical Institute Potsdam, An der Sternwarte 16, 14482 
Potsdam, Germany\\
$^4$ Universidad Aut\'onoma de Madrid, Grupo de Astrofisica, 28049 
Madrid, Spain\\
$^5$ Racah Institute of Physics, Hebrew University, Jerusalem 91904, 
Israel\\
$^{6}$ Stockholm University, SE-106 91 Stockholm, Sweden}
\date{\today} \pubyear{2008} \volume{000}
\pagerange{1} \twocolumn \maketitle
\label{firstpage}

\begin{abstract}
We present the first detailed structure formation and radiative transfer
simulations of the reionization history of our cosmic neighbourhood. To
this end, we follow the formation of the Local Group of galaxies and 
nearby clusters by means of constrained simulations, which use the 
available observational constraints to construct a representation of 
those structures which reproduces their actual positions and properties 
at the present time. We find that the reionization history of the Local 
Group is strongly dependent on the assumed photon production 
efficiencies of the ionizing sources, which are still poorly constrained. 
If sources are relatively efficient, i.e. the process is 'photon-rich', 
the Local Group is primarily ionized externally by the nearby clusters. 
Alternatively, if the sources are inefficient, i.e. reionization is 
'photon-poor' the Local Group evolves largely isolated and reionizes 
itself. The mode of reionization, external vs. internal, has important 
implications for the evolution of our neighbourhood, in terms of e.g. 
its satellite galaxy populations and primordial stellar populations. 
This therefore provides an important avenue for understanding the young 
universe by detailed studies of our nearby structures.   
\end{abstract}


\section{Introduction}
Approximately thirteen billion years ago the cosmic neighbourhood 
destined eventually to become our Local Group of galaxies underwent
a dramatic transition: a giant ionization front swept through,
engulfing it in a sea of ionizing radiation. This occurred as a 
local manifestation of a global transition of the intergalactic 
medium in the whole universe referred to as Cosmic Reionization, 
caused by the radiation from the first galaxies. This process 
converted the intergalactic medium from neutral and cold gas 
during the Cosmic Dark Ages before any galaxies existed, into a 
hot, ionized plasma. 

The absorption spectra of QSOs from redshift 0 to about 6 show
that the intergalactic medium has been almost fully ionized for 
most of the lifetime of the Universe. On the other hand, the 
recent data from the Wilkinson Microwave Anisotropy Probe (WMAP) 
satellite yielded a rather large optical depth for scattering the 
Cosmic Microwave Background photons on free electrons. This 
strongly suggests that the reionization epoch started well before
redshift 10 and therefore was fairly extended in time. It also 
confirmed independently the existence of a reionization epoch, 
required by this additional optical depth. The process of 
reionization had far-reaching consequences for subsequent galaxy 
formation. The photoionization heating which accompanies 
reionization increased the gas temperature from the very low one, 
of order a few K or less, during the Cosmic Dark Ages before the 
first stars formed, to $\sim10^4$~K or more. This in turn increased 
the corresponding Jeans mass, the mass above which gas pressure 
cannot sucessfully counteract gravity, by about 5 orders of 
magnitude. This strongly suppressed the formation of low-mass 
galaxies and cut off the star formation in previously-formed ones, 
and thereby should have significantly influenced the early population 
of dwarf satellite galaxies. For this reason, reionization is often 
invoked as a plausible explanation for the observed lack of galaxy 
satellites compared to the numbers predicted by pure dark matter 
simulations \citep{2000ApJ...539..517B, 2009MNRAS.400.1593M,
2010MNRAS.402.1995M,2010ApJ...710..408B}. The same ionization and 
heating process also changed significantly the character of star 
formation, as stars whose formation starts out from hot, ionized 
gas, even those with a primordial element abundance, are different 
from the ones forming out of initially cold gas 
\citep[e.g.][]{2003ApJ...586....1M}. 

In recent years a variety of theoretical and numerical modelling 
work has shown that reionization was highly patchy in nature, with 
a very large variations in the times at which different regions 
become reionized \citep[e.g.][]{2006MNRAS.369.1625I,
2007MNRAS.376..534I,2007ApJ...654...12Z}. Many of the effects of 
reionization on the later structure formation are highly dependent 
on the stage at which the forming structures are at the time when 
they become reionized. Therefore, it is important to follow the 
reionization history of different patches in detail and to compare 
the timing of reionization with the stage of formation of the local 
structures.

\citet{2007MNRAS.381..367W} used large-scale reionization simulations 
to study the reionization history of galaxies of a variety of 
present-day types: field galaxies, cD central cluster galaxies, 
$L_*$ galaxies and Local Group-like systems. In particular, they
focused on the question which galaxies are reionized internally 
(i.e. by their own progenitors) vs. externally (i.e. before 
sigificant fraction of their mass is in collapsed objects). For 
most galaxy types the answer is statistical, with certain 
probabilities for each outcome. They found that there is a halo
mass scale, of order $10^{12}M\odot$, which divides the massive
galaxies which are predominantly internally reionized and 
lower-mass ones for which the opposite is true. More recently, 
\citet{2009ApJ...703L.167A} studied the same problem
using larger-scale, coarsely-resolved semi-analytical reionization 
calculations. They achieved better statistics for the 
larger halos, from Milky Way-sized up to galaxy clusters, as a
consequence of the larger volume their calculation followed and 
reached very similar conclusions to \citet{2007MNRAS.381..367W}. 
 
However, these previous studies are statistical and therefore can 
only yield a certain probability for a given type of system to be 
externally or internally reionized. In contrast, the answer for a 
specific system, like our own Local Group of galaxies, consisting 
of the Milky Way, Andromeda, M33 and their satellite galaxies, will 
depend on the details of its and its neighbouring systems' formation, 
the timing of that formation, and the relative positions in space of 
their projenitors at the relevant epochs. This kind of detailed 
information can only be obtained through numerical simulations 
with constrained initial conditions. 

Constrained simulations aim to reproduce the spatial and velocity 
structure of our Local Group and its neighbourhood at present 
(redshift $z=0$). In this work we combine the most advanced constrained 
realizations of the Local Group available at present with an accurate 
treatment of the radiative transfer during cosmic reionization. This 
allowed us for the first time to calculate the specific reionization 
history of all Local Group progenitors, as well as those of the nearby 
structures like the Virgo and Fornax clusters of galaxies.    

\section{Methodology}

\subsection{Constrained simulations of the local universe}

The optimal way of constructing a numerical simulation that closely 
reproduces our local cosmological neighborhood is provided by the 
\citet{1991ApJ...380L...5H} algorithm for making a constrained 
realizations of Gaussian random field. Given our ability to extract 
observational data that can be imposed as linear constraints on the 
primordial perturbation field, this method can be used to construct 
initial conditions that obey these constraints. Here we follow 
\citet{2003ApJ...596...19K} and impose two types of data on the 
simulation. The first data  set consists of peculiar velocities of 
galaxies, drawn from the MARK III \citep{1997ApJS..109..333W},
surface brightness fluctuations \citep{2001ApJ...546..681T} 
catalogues and the Catalog of Nearby Galaxies 
\citep{2004AJ....127.2031K}. The other data set is obtained from 
the catalog of nearby X-ray selected clusters of galaxies 
\citep{2002ApJ...567..716R}. The main obstacle faced here is how to 
translate the present epoch observables into quantities that are 
linear in the primordial perturbation fields. Peculiar velocities 
evolve more slowly than the density field and are assumed here to 
be linear. The present epoch virial parameters of the clusters are 
processed by the spherical top-hat model to produce their linear 
overdensity. The data used here and its associated observational 
errors effectively constrain the LSS on scales larger than $\approx 
5 \hmpc$ \citep[cf.][]{2003ApJ...596...19K}. The simulation used 
here is designed to reproduce the Local Supercluster, harbouring a 
Virgo-like cluster. Such a configuration is easily reproduced by 
the constrained simulations. Local Group (LG) like objects, on the 
other hand, are randomly emerging in the simulations. The simulations 
used here were each selected out of a few to have a LG-like object, 
similar to the observed one. They are described in more detail in
\cite{clues}. 

Starting from the above constrained initial conditions, the GADGET-2 
code \citep{2005MNRAS.364.1105S} was used to follow the dark matter 
in a $L=64 \hmpc$ computational box, spanned by $1024^3$ particles,
starting at redshift $z=100$. The cosmological parameters given by 
the WMAP 3-year data have been adopted ($\Omega_M = 0.24, 
\Omega_\Lambda=0.76, h=0.73, \Omega_b=0.0418, \sigma_8 =0.75, n=0.95$), 
giving a particle mass of $m_{\rm DM} = 1.63 \times 10^7 \hmsun$. (A more 
detailed description is provided in \citet{2009arXiv0906.0585Z}).

\begin{figure}
\begin{center}  
\includegraphics[width=3.4in]{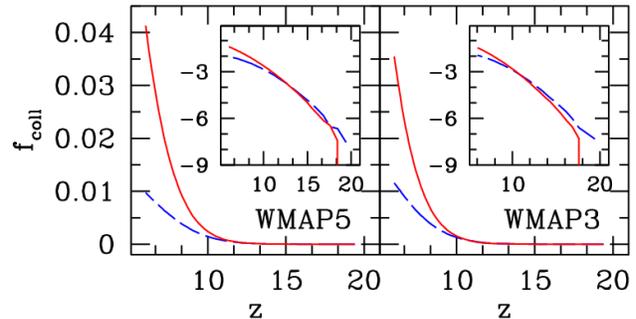} 
\caption{
Collapsed fraction in high-mass ($M>10^9\,M_\odot$; red, solid) and 
low-mass sources ($M<10^9\,M_\odot$; blue, dashed) (insets show the 
same in log scale) vs. cosmic redshift for WMAP5 (left) and WMAP3 
(right) cases.
\label{fcoll_sourcenum_fig}}
\vspace{-.3in}
\end{center}
\end{figure}

\begin{figure}
\begin{center}  
\includegraphics[width=3.2in]{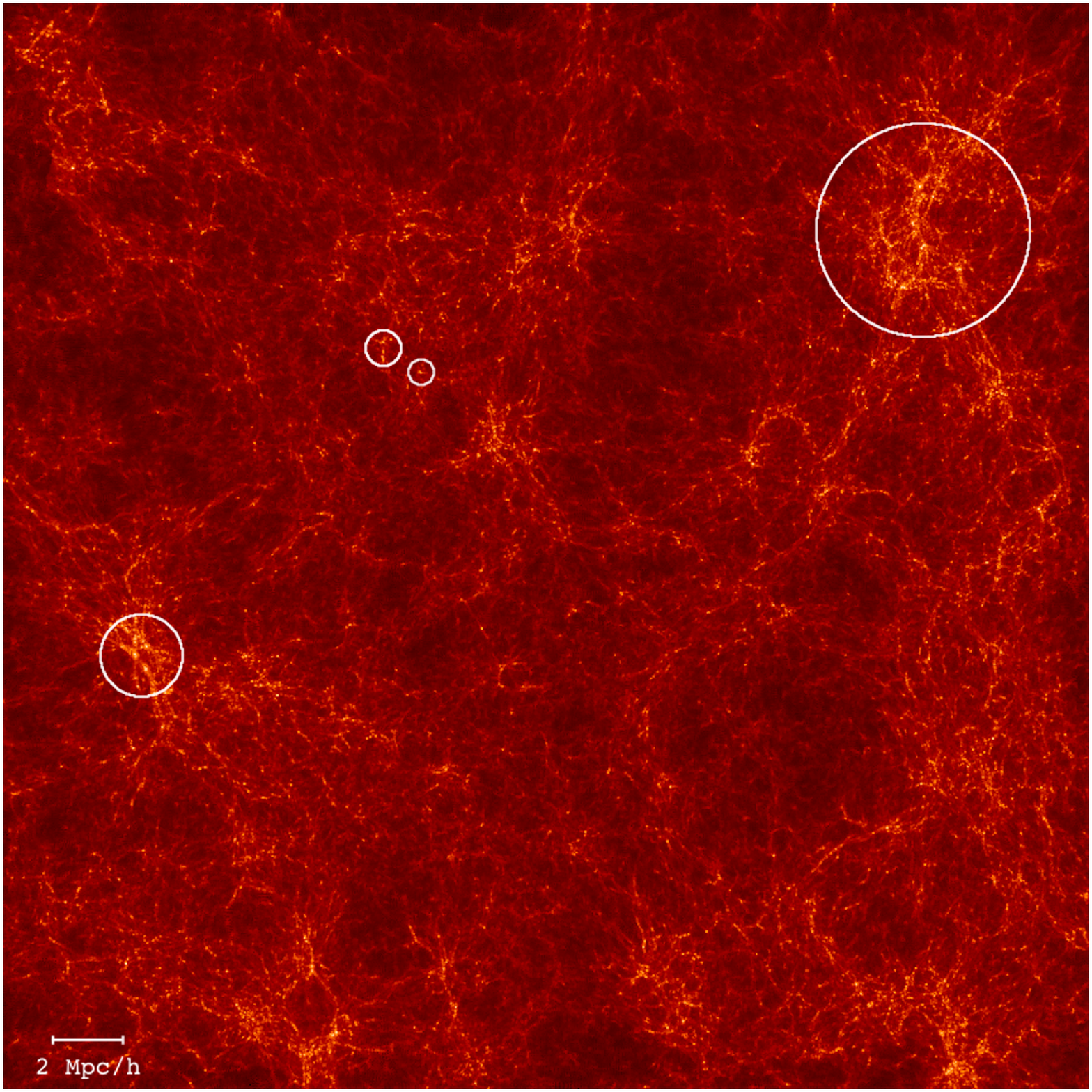}
\caption{Matter distribution in a volume of $30 \hMpc \times 30 \hMpc 
\times 7 \hMpc$ depth in comoving Mpc at redshift $z = 9$. The circles 
with decreasing radius according to the decreasing mass of the objects 
denote the regions where the progenitors of Virgo, Fornax, M31 and 
MW are situated. The slice is situated in the supergalactic YZ-plane.
\label{dens_slice_fig}}
\end{center}
\end{figure}

In order to check the robustness of our results with respect to the 
particular realization of the constrained simulation, we also performed 
a second simulation with an independent underlying random realization. 
This second simulation implements the same constraints on the 
local structures at the present time. 
The simulation volume and number 
of particles are the same as above, while the background cosmology is 
now based on the WMAP 5-year data, combined with the available constraints 
from the large-scale structure (BAO) and supernovae 
 ($\Omega_M = 0.279, \Omega_\Lambda=0.721, h=0.73, \Omega_b=0.046, 
\sigma_8 =0.817, n=0.96$).
In this case we include in the analysis not just the 
Local Group and Virgo, but also our other nearby galaxy cluster, Fornax. 
In Figure~\ref{dens_slice_fig} we show the local density distribution
at redshift $z=9$ within a slice of comoving $30 \hMpc \times 30 \hMpc $ 
and $ 7 \hMpc$ depth which contains the projenitors of the objects of 
interest, Virgo, Fornax, M31 and Milky Way. The slice is situated in the 
supergalactic YZ-plane. The main progenitors of these 
objects have at redshift $z=9$ masses of 
$1.2 \times 10^{11} \hMsun$, 
$6.6 \times 10^{10} \hMsun$, 
$3.8 \times 10^{9} \hMsun$, and 
$3.4 \times 10^{9} \hMsun$, respectively.
The marked Virgo (Fornax) regions are populated with about 2000 (700)
cluster progenitors, each with a mass larger than $4 \times 10^{8} \hMsun$.
This leads to a total progenitor mass of $2.4 \times 10^{12} \hMsun$
($0.9 \times 10^{12} \hMsun$). One can clearly see the low density
regions between the progenitors of the clusters and the progenitors of
the Local Group. There are a few other objects next to the Local Group
but with much smaller masses than that concentrated in the
proto-cluster regions.

All our simulations were performed within the CLUES (Constrained Local 
UniversE Simulations) project\footnote{http://clues-project.org}.

\subsection{Radiative transfer simulations}

The radiative transfer simulations were performed using a radiative 
transfer and non-equilibrium chemistry code called C$^2$-Ray 
\citep{methodpaper}, tested in detail as discussed in 
\citet{methodpaper}, \citet{comparison1} and  \citet{comparison2}. 
Our simulation methodology was presented in \citet{2006MNRAS.369.1625I,
21cmreionpaper} and \citet{2007MNRAS.376..534I}. The underlying N-body 
structure formation simulations discussed in the previous section 
provide a time sequence of density distributions and catalogues of 
identified halos. Gas is assumed to follow the dark matter 
distribution, which at these scales ($0.25\,h^{-1}$Mpc$ - $64$\,h^{-1}$Mpc)
is a very good approximation. We have produced 53 density slices and 
halo catalogues roughly equally spaced in time between the redshifts 
20 and 6, every $\sim14.5$~Myr. All halos are assumed to host galaxies 
and thus to be potential 
sources of ionizing radiation. Each is assigned an ionizing 
luminosity proportional to its mass. Low-mass sources, with total 
mass below $10^9\,M_\odot$, are assumed to be active in the neutral 
regions, but to be suppressed in the ionized regions. Therefore, 
once their cell is ionized their emissivity is set to zero. These 
assumptions are based on the fact that the Jeans mass rises 
significantly when a region is ionized and heated, which thereby 
limits the fresh gas infall onto halos and supresses the future 
formation of low-mass galaxies, see \citet{2007MNRAS.376..534I} 
for a more detailed discussion.

Both the density and halos are binned on a $256^3$ grid for 
the radiative transfer processing. Halos in the same RT cell 
are combined and a luminosity is assigned following a simple, 
physically-motivated prescription. The emissivity of the 
ionizing sources - how many ionizing photons produced in 
galaxies over some time period reach the intergalactic medium 
- depends on what fraction of the galactic gas is converted 
into stars, how effective are the stars at producing ionizing 
photons and, finally, what fraction of the photons manage to 
escape the galaxy. We parametrise it with a single parameter 
$f_\gamma$, which is equal to the number of ionizing photons per 
atom in galaxies which reach the intergalactic medium in the time 
between two consecutive time slices\footnote{One can also introduce
a slightly different efficiency parameter, $g_\gamma$, given by
$g_\gamma=f_\gamma\left(\frac{10 \;\mathrm{Myr}}{\Delta t}\right)\,$
where ${\Delta t}$ is the time between two snapshots from the N-body
simulation. This has the advantage that it is a rate per unit time 
and as such it is independent of ${\Delta t}$, which makes easier 
comparisons between simulations with different ${\Delta t}$.} The 
main difference between our two adopted background cosmologies is 
in the density fluctuation normalization ($\sigma_8$). A higher 
$\sigma_8$ yields higher halo collapsed fractions at any given epoch 
and a correspondingly larger number of collapsed halos, as illustrated 
in Figure~\ref{fcoll_sourcenum_fig}. In terms of reionization this 
means that for a fixed ionizing emissivity proportional to the
collapsed fraction the evolution is shifted to somewhat earlier times, 
resulting in an earlier overlap epoch \citep{2006ApJ...644L.101A}.
This effect was much larger for WMAP 1-year ($\sigma_8=0.9$) vs. WMAP
3-year data ($\sigma_8=0.74$) than it is for WMAP 5-year data 
($\sigma_8=0.8$) vs. the WMAP 3-year data. 

Much larger uncertainty concerns the properties of high redshift 
sources. Because of the still scarse observational data those are 
not at well constrained at present and hence $f_\gamma$ is largely 
a free parameter within certain, fairly wide, bounds. Furthermore, 
the radiative transfer simulations are quite computationally 
expensive, and therefore it is not practical to investigate the 
full available parameter space. We address this problem by 
considering two representative cases which are roughly bracketing 
the range of expected behaviour, while at the same time satisfying 
the available global observational constraints, as follows. For our 
Model 1 simulation we use the first N-body realization, with WMAP3 
cosmology and we adopt $f_\gamma=100$ for the massive sources 
($M>10^9\,M_\odot$) and $f_\gamma=250$ for the low-mass, suppressible 
ones ($M<10^9\,M_\odot$). These relatively high efficiencies yield a 
fairly fast, photon-rich reionization process with an early overlap. 
For our second simulation, on the other hand, we use the second 
constrained realization, with WMAP5 cosmology, and we adopt a lower 
source efficiencies of $f_\gamma=10$ for the massive sources and 
$f_\gamma=150$ for the low-mass ones. This yields a more extended, 
photon-poor reionization history. For conciseness of notation, we 
will to refer to the two combinations of N-body realizations and 
corresponding radiative transfer simulation simply as the Model 1 and 
Model 2 cases, with the implicit understanding that this implies a
different constrained realization and assumed photon emissivities of 
the sources, as well as different background cosmology.

\section{Results} 

\subsection{Global reionization history}
\begin{figure}
\begin{center}  
\includegraphics[width=3.2in]{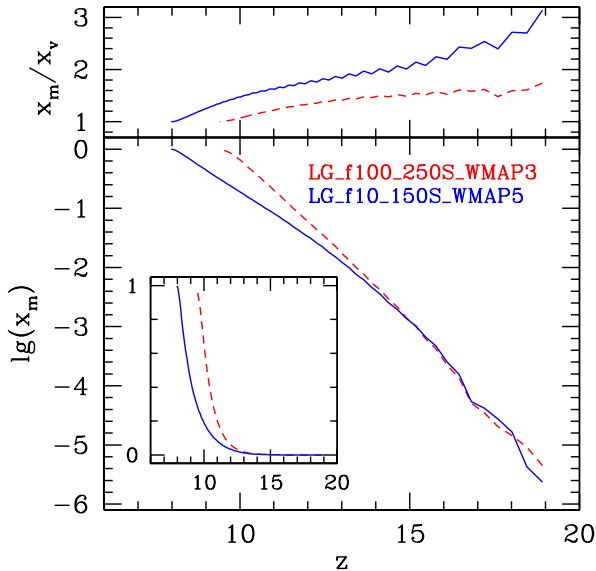} 
\caption{(Bottom) Evolution of the mass-weighted ionized fractions, 
$x_m$ for Model 1 (red, dashed) and Model 2 (blue, solid) cases, 
inset shows the same in linear scale; and (top) the corresponding
ratios of mass-weighted to volume-weighted ionized fractions, which 
corresponds to the average density of the ionized regions in units 
of the mean, versus redshift z.
\label{global_reion_hist_fig}}
\end{center}
\end{figure}
The global mass-weighted reionization histories produced by our two 
simulations are shown in Figure~\ref{global_reion_hist_fig}. In 
either case reionization starts when the first resolved halos (which
correspond to the first ionizing sources) form, around $z\sim20$. 

\begin{figure*}
\begin{center}  
\includegraphics[width=3.2in]{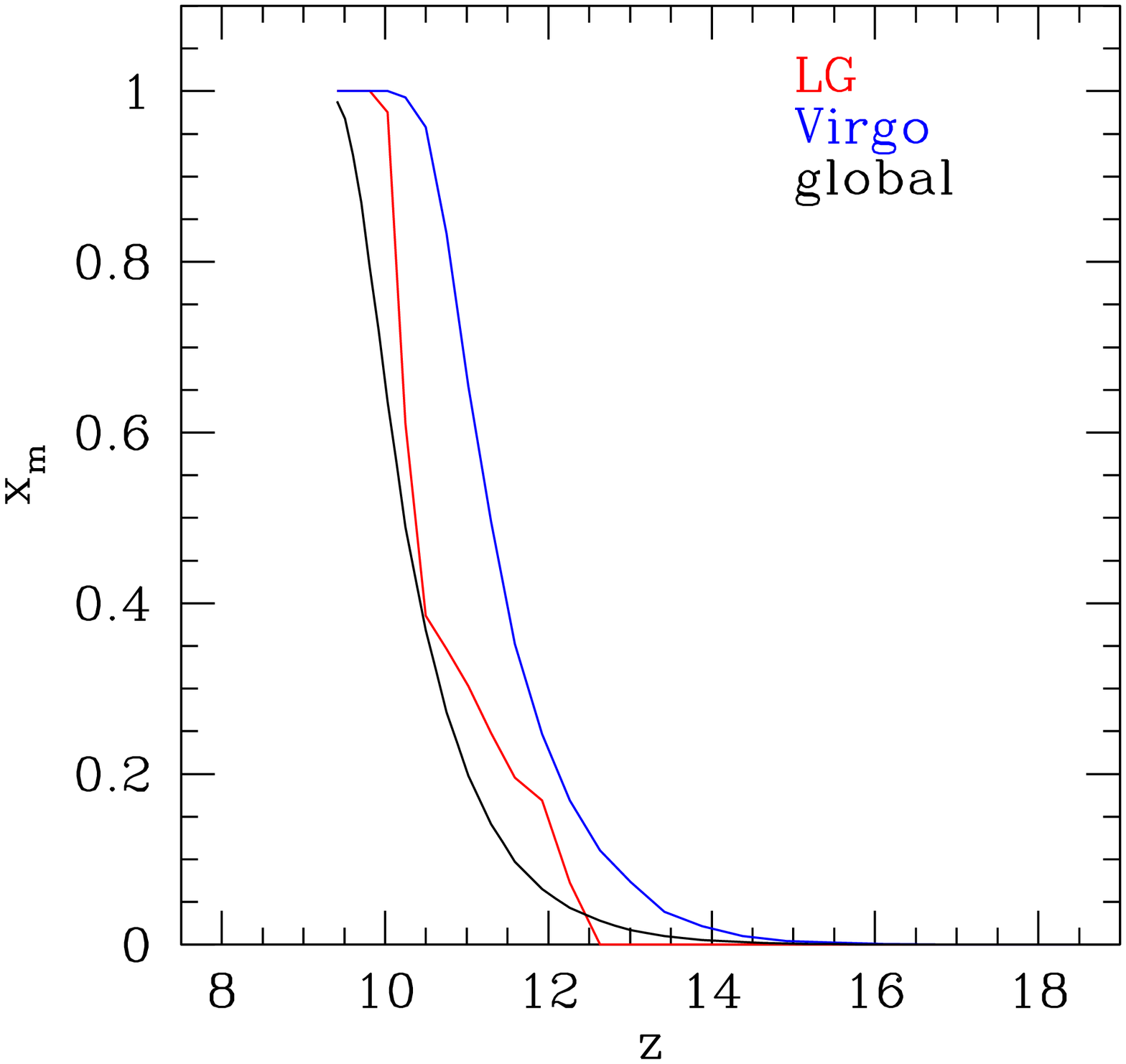} 
\includegraphics[width=3.2in]{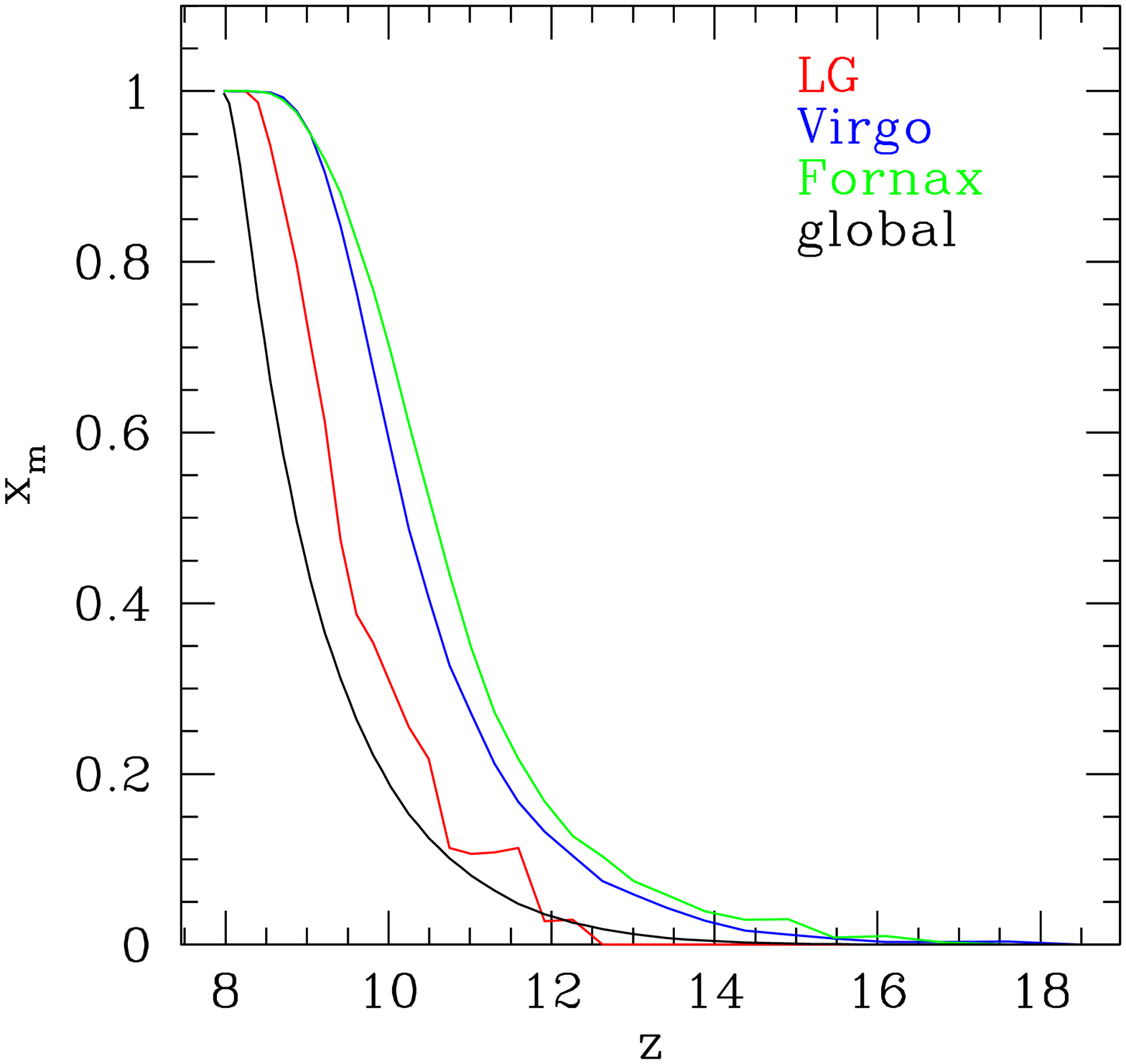} 
\caption{Mass-weighted mean ionized fractions, $x_m$, 
for the Local Group, nearby clusters and global mean (as indicated 
by color) vs. redshift for Model 1 (left; left to right: global, LG, 
Virgo) and Model 2 (right; left to right: global, LG, Virgo, Fornax). 
\label{reion_hist_fig}}
\end{center}
\end{figure*}
Our adopted $f_\gamma$ values yield final overlap of ionized regions at 
$z_{\rm ov}\sim9$ (8) for the Model 1 (Model 2) case, in rough agreement 
with the current observational constraints. The corresponding integrated 
Thomson scattering optical depth seen by the Cosmic Microwave Background 
photons is $\tau_{\rm es}=0.094 (0.069)$ is also in agreement with the 
latest constraints from WMAP satellite combined with the other available
datasets, $\tau_{\rm es}=0.084\pm0.016$ \citep{2009ApJS..180..330K}.

The early reionization ($z>14$) is driven primarily by the low-mass 
sources, which have similar efficiencies in the two cases (the 
slightly lower source efficiency in the Model 2 case is compensated 
for by its higher collapsed fraction at the same redshift) and as a 
consequence the two reionization histories are initially very similar. 
Later on the larger sources take over, both because of their rapidly 
rising collapsed fraction (cf. Figure~\ref{fcoll_sourcenum_fig})
and the strong Jeans suppression of the low-mass sources, and thus 
reionization proceeds more slowly in the Model 2 case due to the lower 
efficiency adopted for its high-mass sources. The fact that the mean
mass-weighted ionized fraction, $x_m$ is always larger than the 
corresponding volume-weighted one, $x_v$ 
(Figure~\ref{global_reion_hist_fig}, upper panel) 
indicates that reionization proceeds in an inside-out fashion (i.e. 
high-density regions are preferentially ionized first) in both cases, 
in agreement with previous simulation results based on non-constrained 
realizations \citep{2006MNRAS.369.1625I}. For the Model 2 realization 
this ratio is noticeably higher, up to $\sim3$ at $z=20$, due to its 
more advanced structure formation at any given redshift. This also 
boosts the clumpiness of the gas, and therefore the recombinations, 
which in turn extends reionization even further.

\subsection{Local reionization histories}

The mean reionization history presented above ensures that the 
currently available global observational constraints - the 
electron-scattering optical depth and overlap epoch are satisfied. 
However, to achieve our present goals we need to track separately 
the reionization history of the progenitors of each object of 
interest, namely the Local Group, as well as the nearby clusters 
of galaxies. To this purpose, we extracted the Lagrangian mass 
distribution for each object (i.e. the mass which eventually will 
end up in that object by the present day) and followed the 
reionization history of all radiative transfer cells containing 
at least one particle which ends up in that object by $z=0$. The 
resulting local reionization histories are shown in 
Figure~\ref{reion_hist_fig}, along with the global one for direct 
comparison. 

\begin{figure*}
\begin{center}  
\includegraphics[width=3.2in]{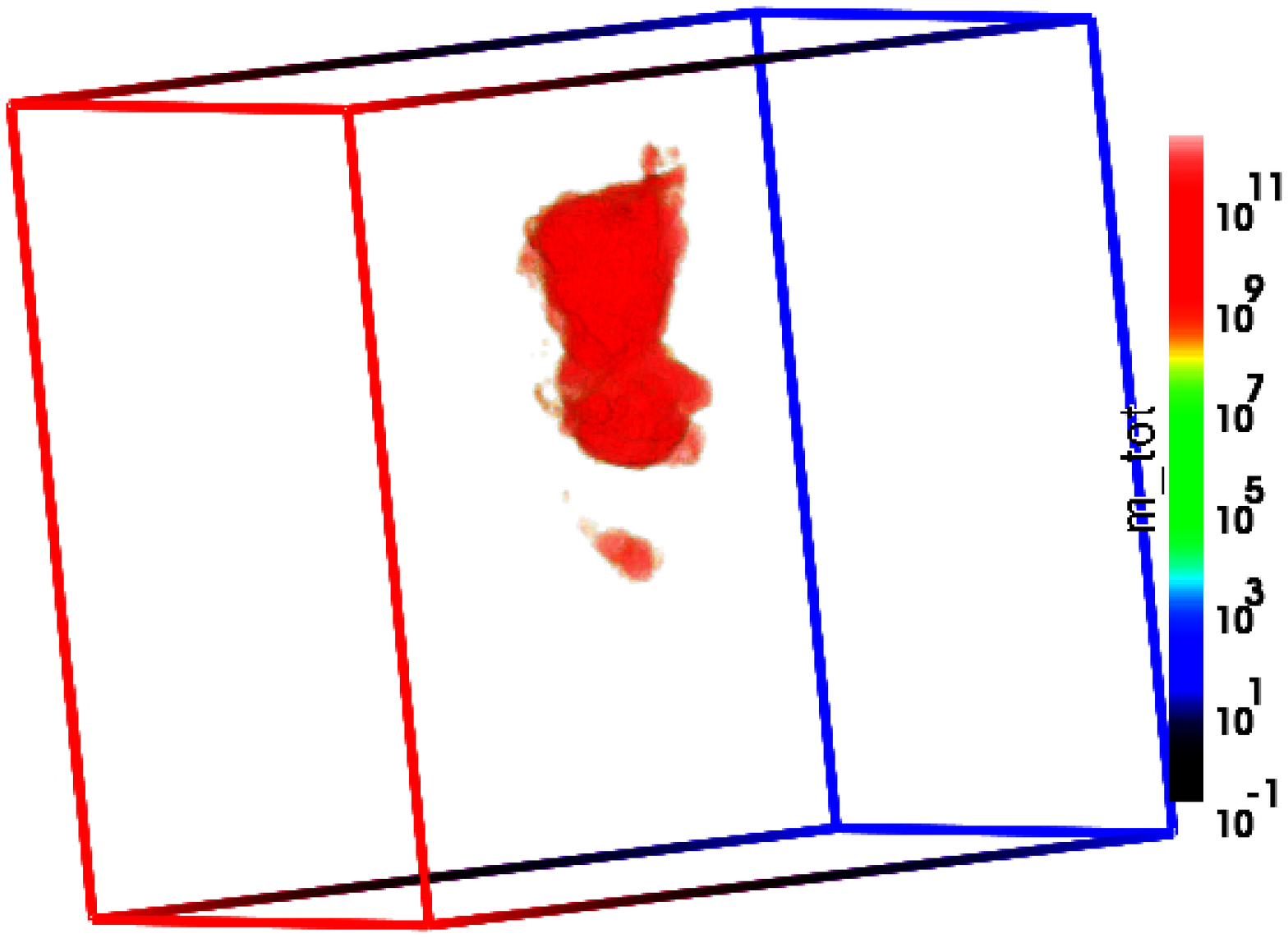} 
\includegraphics[width=3.2in]{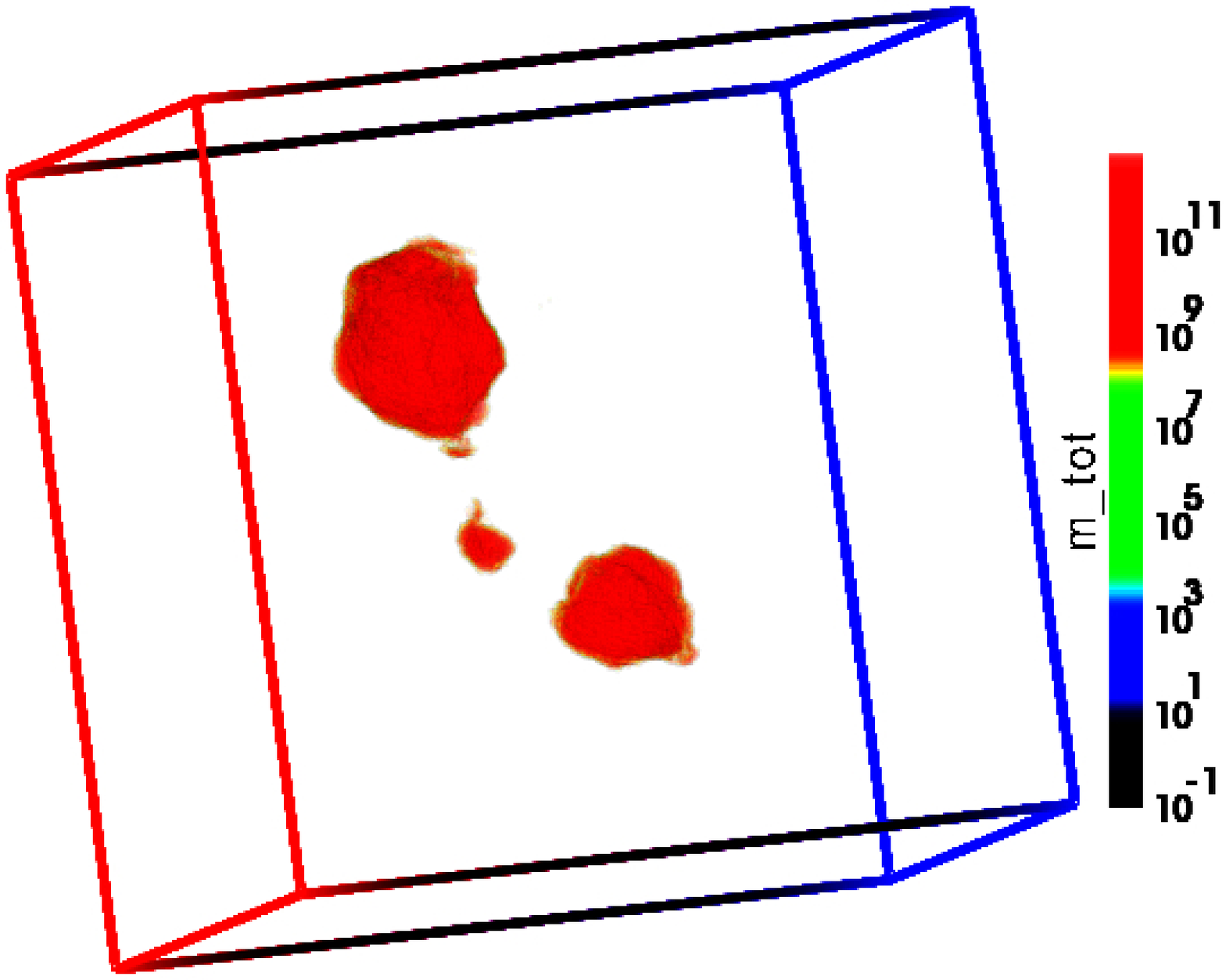}
\caption{Volume renderig of the total mass, which by the present time 
($z=0$) will end up as part of the objects of interest here: (left) 
Virgo cluster (top object) and the Local Group (bottom object) at 
redshift $z=10$; (right) Virgo cluster (top object), the Local Group 
(middle, smaller object) and Fornax (bottom object) at redshift $z=9$. 
The color scale units are Solar masses per radiative transfer cell.
\label{3D_tot_mass_fig}}
\end{center}
\end{figure*}
The Lagranian regions of both the LG and the nearby clusters 
are significantly overdense in either constrained realization
and at all times, reflecting the fact that all of these objects 
correspond to high peaks of the density field. The proto-Local 
Group region starts only moderately overdense, by $\sim7\%$ 
($\sim9\%$) in the Model 1 (Model 2) case, which rises over time 
as the corresponding object collapses gravitationally, to reach 
$\sim16\%$ ($\sim25\%$) by the global overlap epoch at $z\sim9$ 
(8). The proto-clusters correspond to still higher peaks of the 
density field. Initially the proto-Virgo region is overdense by 
$12\%$ for both simulations, rising over time to $24\%$ ($30\%$) 
for Model 1 (Model 2). The proto-Fornax region (Model 2) starts 
$10\%$ overdense, rising to $26\%$ by the global overlap. The 
higher local density yields an (exponentially) larger halo 
collapsed fraction and thus ionizing photon production. Therefore, 
for both LG and clusters we can expect local reionization to occur 
earlier than average, which is confirmed by our simulation results 
(Figure~\ref{reion_hist_fig}).

In both simulations the LG reionization starts at about $z\sim12.5$, 
at which time its oldest progenitor halos form. Before $z\sim12.5$ 
the LG ionized fraction is tiny, below $3\times10^{-9}$ ($2\times10^{-5}$) 
for Model 1 (Model 2). Thereafter the (proto-)LG reionization 
accelerates, albeit only gradually. For Model 1 
(Figure~\ref{reion_hist_fig}, left) the LG ionized fraction 
reaches 17\% by $z\sim12$ and 39\% by $z\sim10.5$. After that 
point the evolution becomes very fast and full ionization 
($x_m>99\%$) is achieved by $z\sim10$. 
In contrast, the reionization history of the (proto-)Virgo cluster 
in the same Model 1 simulation is quite different. Proto-Virgo is 
a higher density peak and the formation of the local nonlinear 
structures is therefore accelerated. Hence the local reionization 
proceeds faster, as well. The evolution remains smooth throughout, 
with no sudden changes of slope, unlike in the LG case. The 
mass-weighted ionized fraction reaches 11\% by $z=12.6$, 50\% by 
$z=11.3$, and 96\% by $z=10.5$.  

The reionization histories are similar in the Model 2 case 
(Figure~\ref{reion_hist_fig}, right). Once again, the proto-cluster
regions, both Virgo and Fornax, reionize earlier than LG and much
ealier than an average region - $x_m=0.1$ is reached by $z=12.3$ 
(12.6), $x_m=0.5$ by $z=10.25$ (10.5) and $x_m=0.9$ by $z=9.2$ (9.3) 
for Virgo (Fornax). For both proto-clusters overlap ($x_m=0.99$) is
reached at $z=8.7$ and the evolution remains smooth throughout. 
Interestingly, most reionization stages (but not the local overlap, 
which is roughly simultaneous) of the Fornax reionization occur 
earlier than the corresponding ones for Virgo, even though Fornax 
has lower mass at the present epoch. In comparison, the reionization 
of the LG occurs later, reaching $x_m=0.1$ by $z=11.6$, $x_m=0.5$ by 
$z=9.4$, $x_m=0.9$ by $z=8.6$ and local overlap is achieved by $z=8.4$. 
It lags the global mean in its earliest stages ($z>12.5$), but as more 
progenitor halos form it catches up and then speeds ahead after $z=12$. 
The Local Group reionization history is again much less smooth than the 
proto-cluster ones, with significant changes of slope around $z\sim12$, 
$z\sim11$ and $z\sim9.6$. Evolution becomes extremely fast after 
$z\sim9.6$, whereby the ionized mass fraction jumps from 0.4 to 1 over 
redshift interval of just $\Delta z\sim1$. Compared to the Model 1 case 
the reionization histories for Model 2, both mean and local are ones, 
are much more extended in time due to the lower source emissivities 
assumed.

\begin{figure*}
\begin{center}  
\includegraphics[width=3.2in]{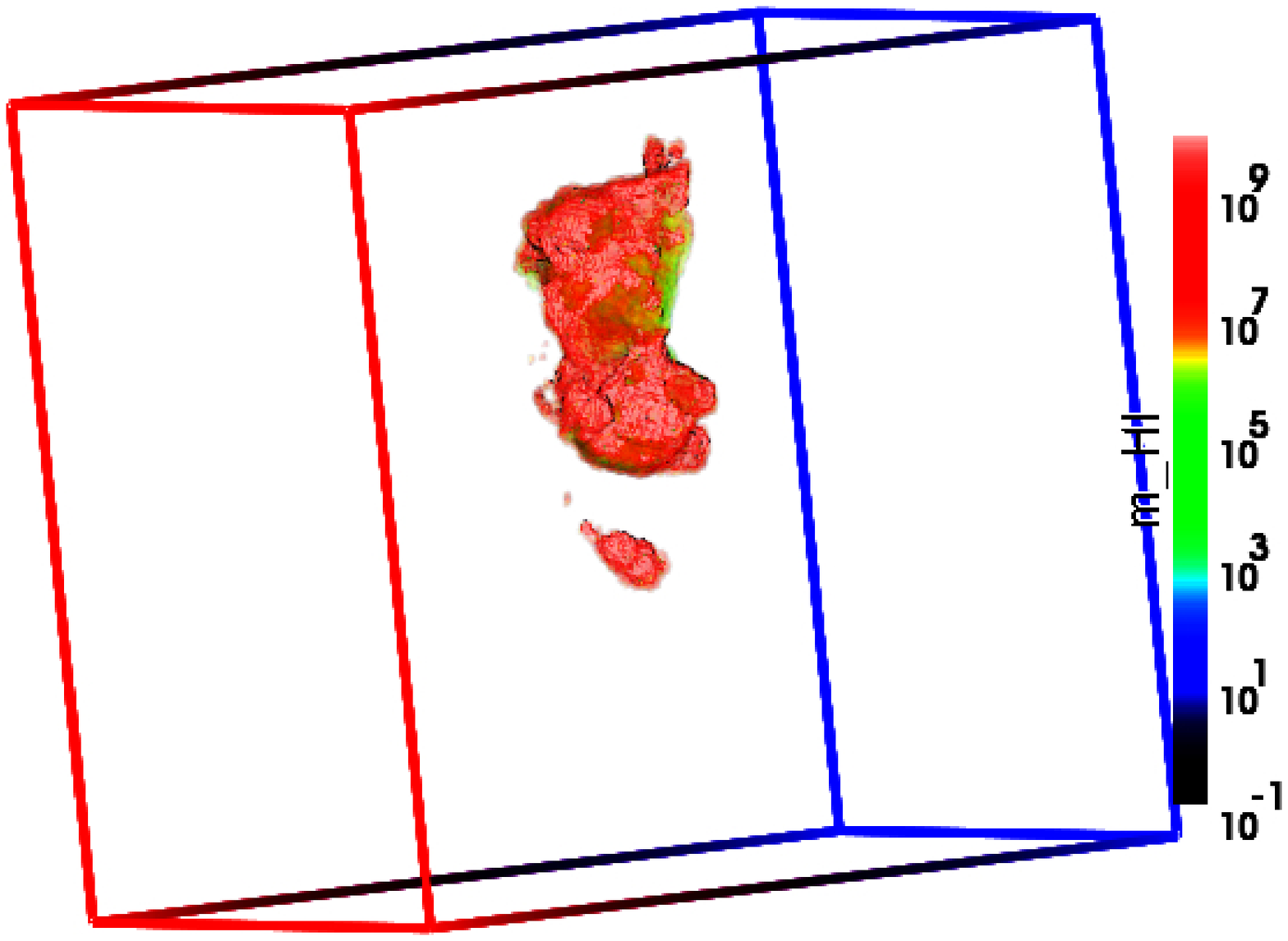} 
  \includegraphics[width=3.2in]{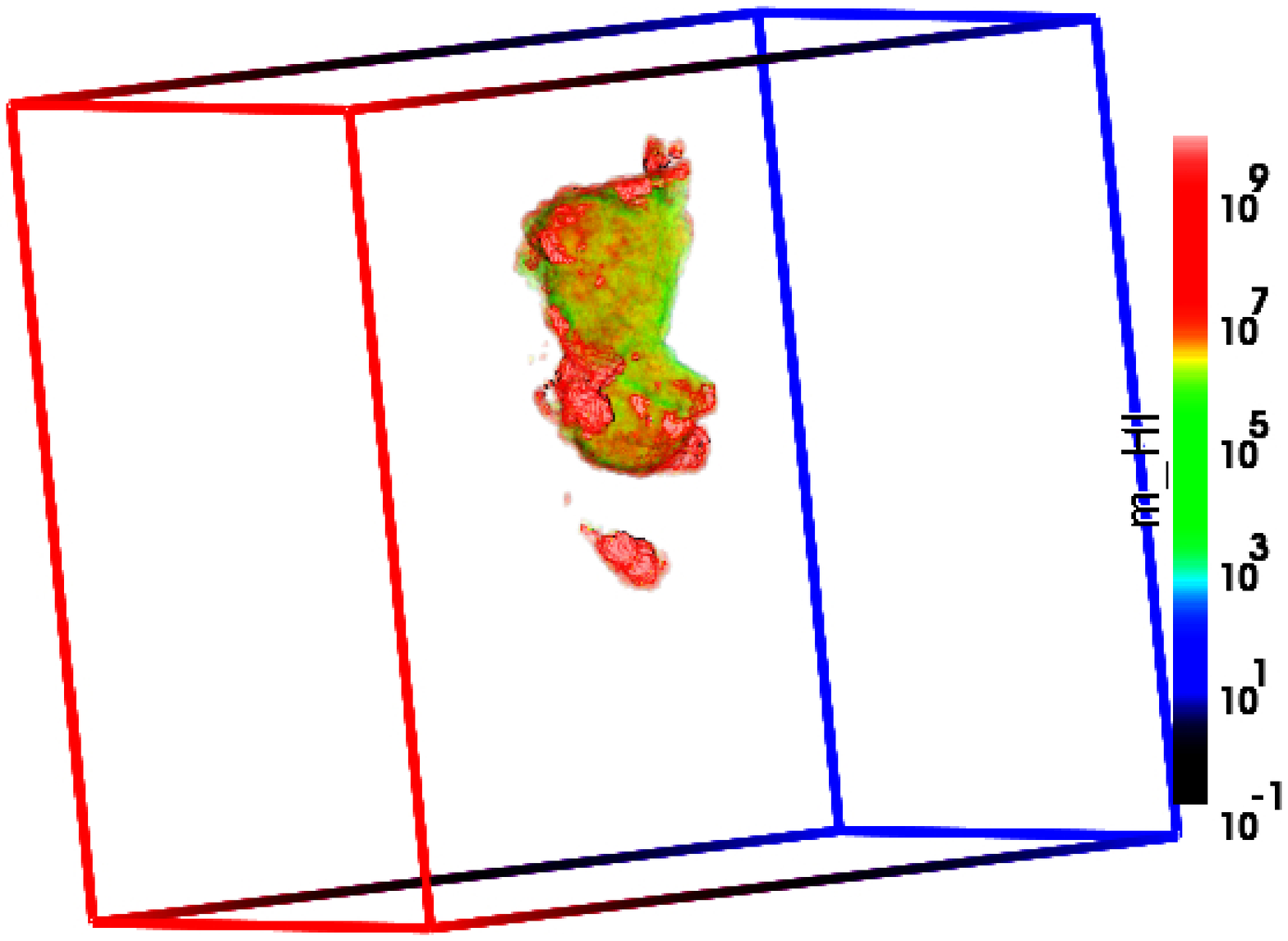}
  \includegraphics[width=3.2in]{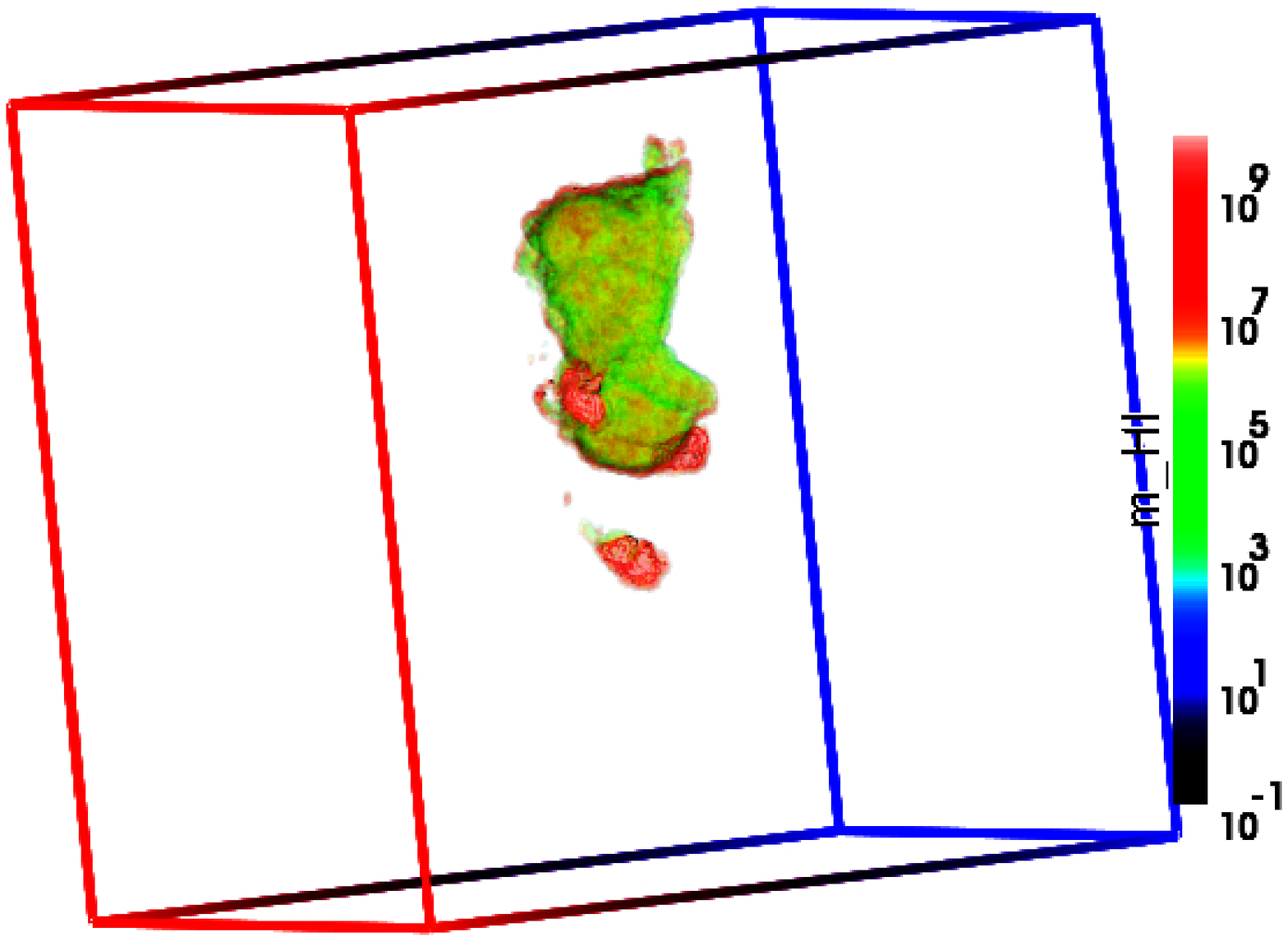}
  \includegraphics[width=3.2in]{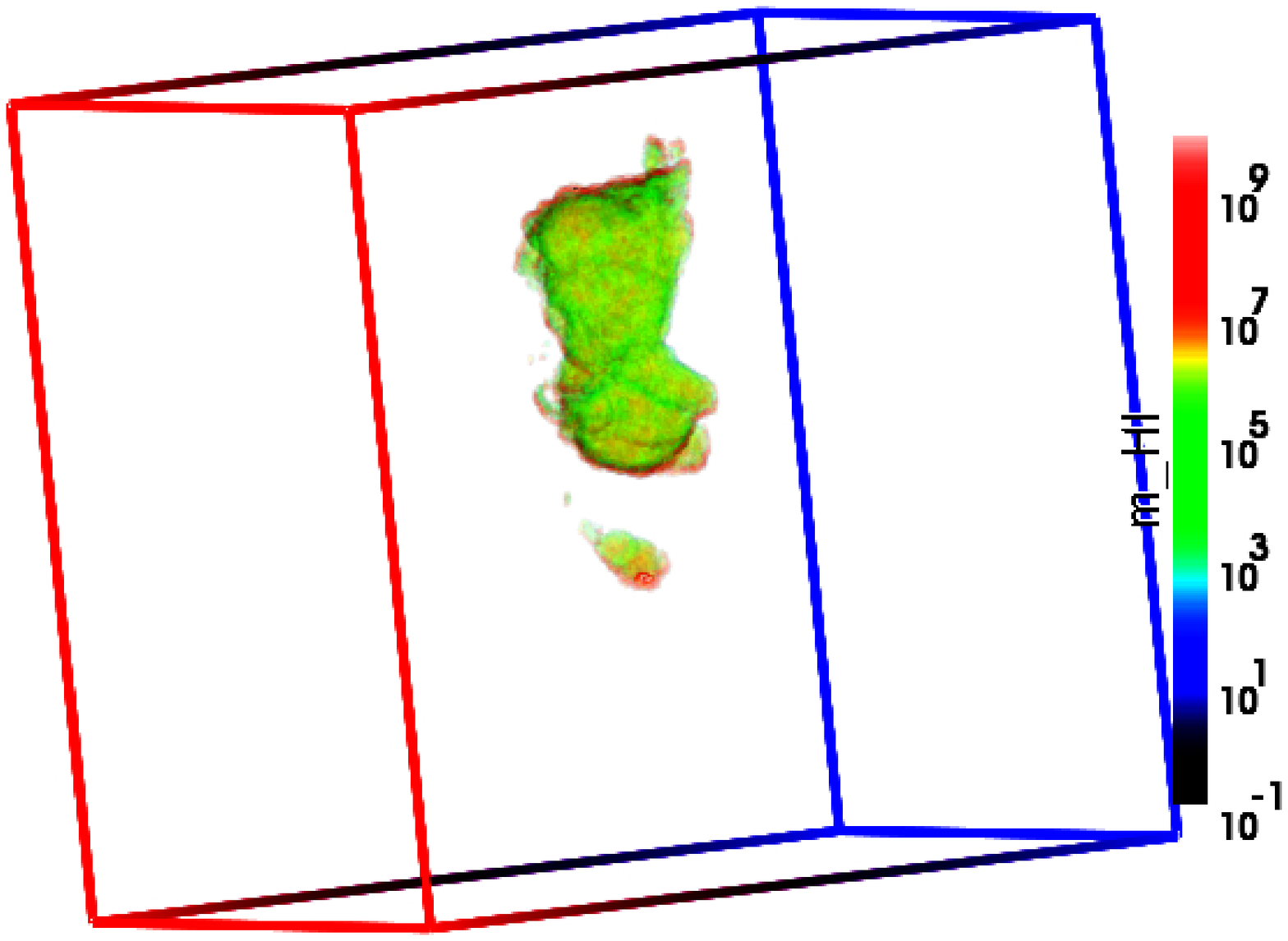}   
\caption{Evolution of the neutral mass at (top to bottom and left 
to right) redshifts $z=10.75,10.5,10.25$ and 10 for Model 1. Red 
is neutral, green is ionized.
\label{3D_HI_fig}}
\end{center}
\end{figure*}

Moving on to a more visual representation of the local reionization 
history, in Figure~\ref{3D_tot_mass_fig} we show a 3D volume rendering 
of the distribution of mass which is destined by the present time 
($z=0$) to become part of the Local Group and nearby clusters for 
Model 1 at $z=10$ (left) and for Model 2 at $z=9$ (right). In the 
realization corresponding to Model 1 the proto-Virgo is a quite large, 
elongated object, extending for about $10\times20$ comoving Mpc, which 
corresponds to about $1$ proper Mpc at this early time, while the 
proto-Local Group is a much smaller object a few Mpc in size, about 
7 comoving Mpc (less than 1 physical Mpc) away from proto-Virgo. In 
the alternative, Model 2 realization the size of the proto-Local 
Group and its distance from Virgo are similar to the Model 1 
realization. However, the proto-Virgo is more compact, less extended 
object and a proto-Fornax is identified, as well, at a similar 
distance from LG, positioned in almost diametrically opposite 
direction. Proto-Fornax is a somewhat less extended object than 
proto-Virgo.

\begin{figure*}
\begin{center}  
\includegraphics[width=2.4in]{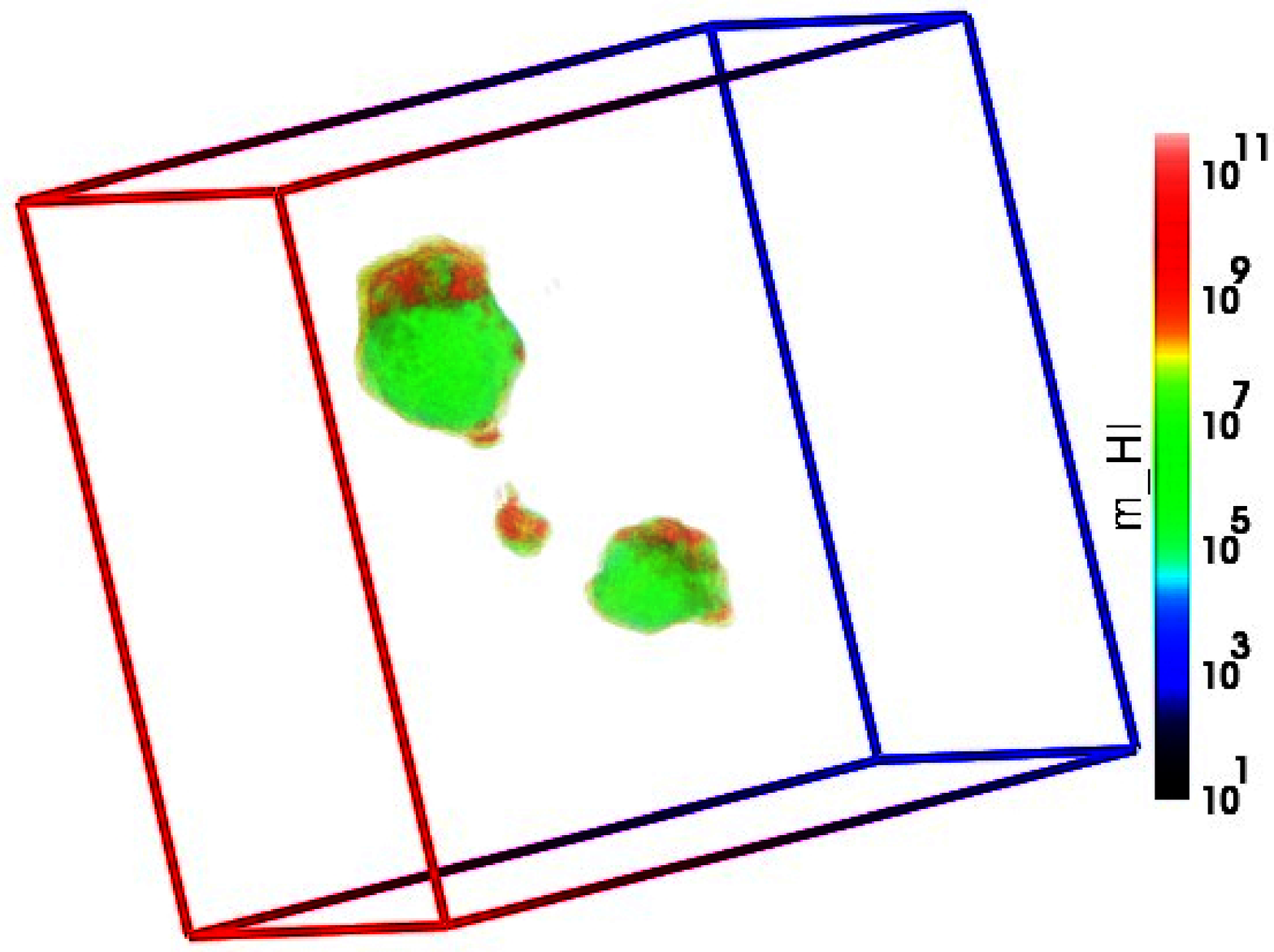} 
\includegraphics[width=2.4in]{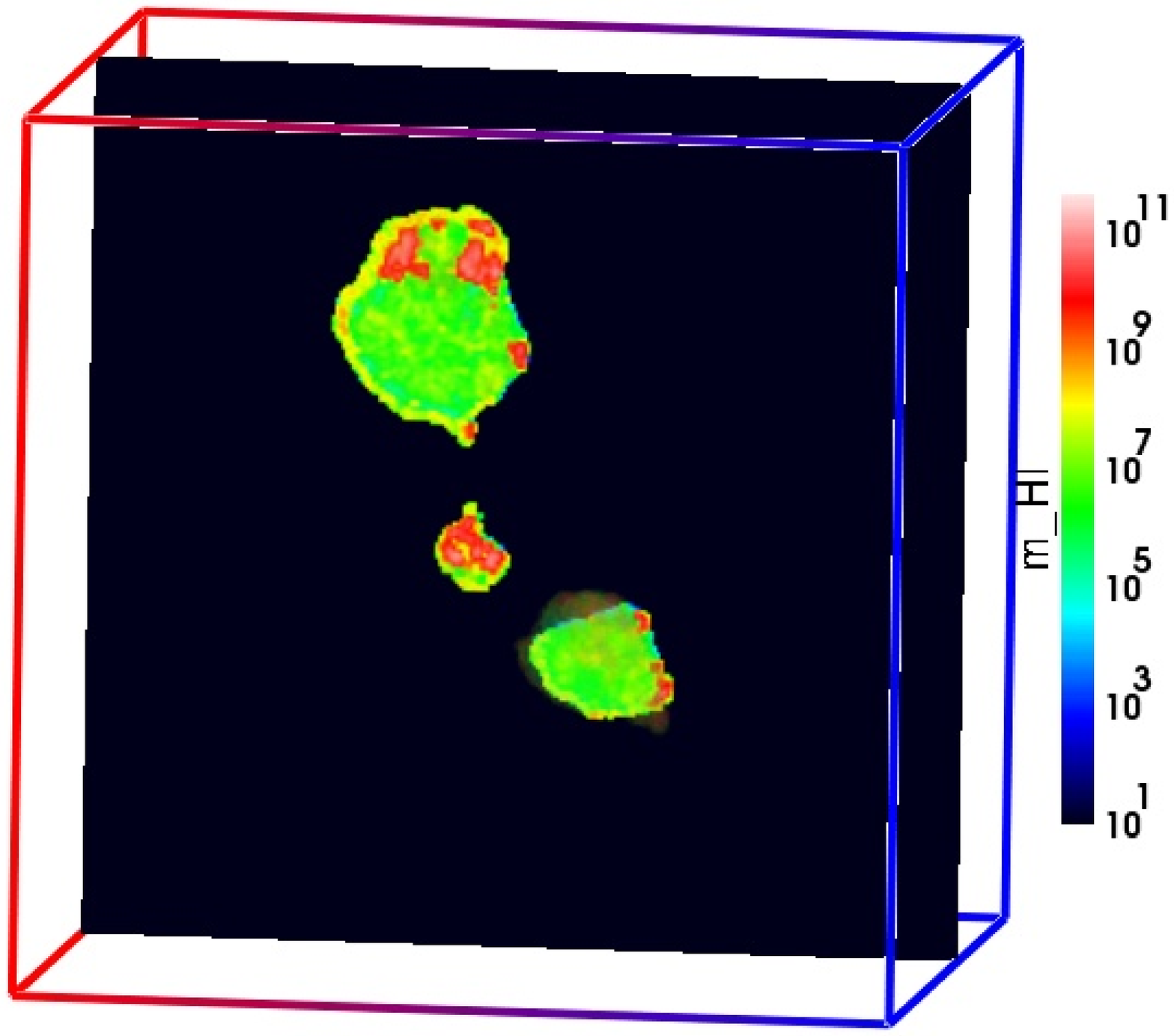} 
\includegraphics[width=2.4in]{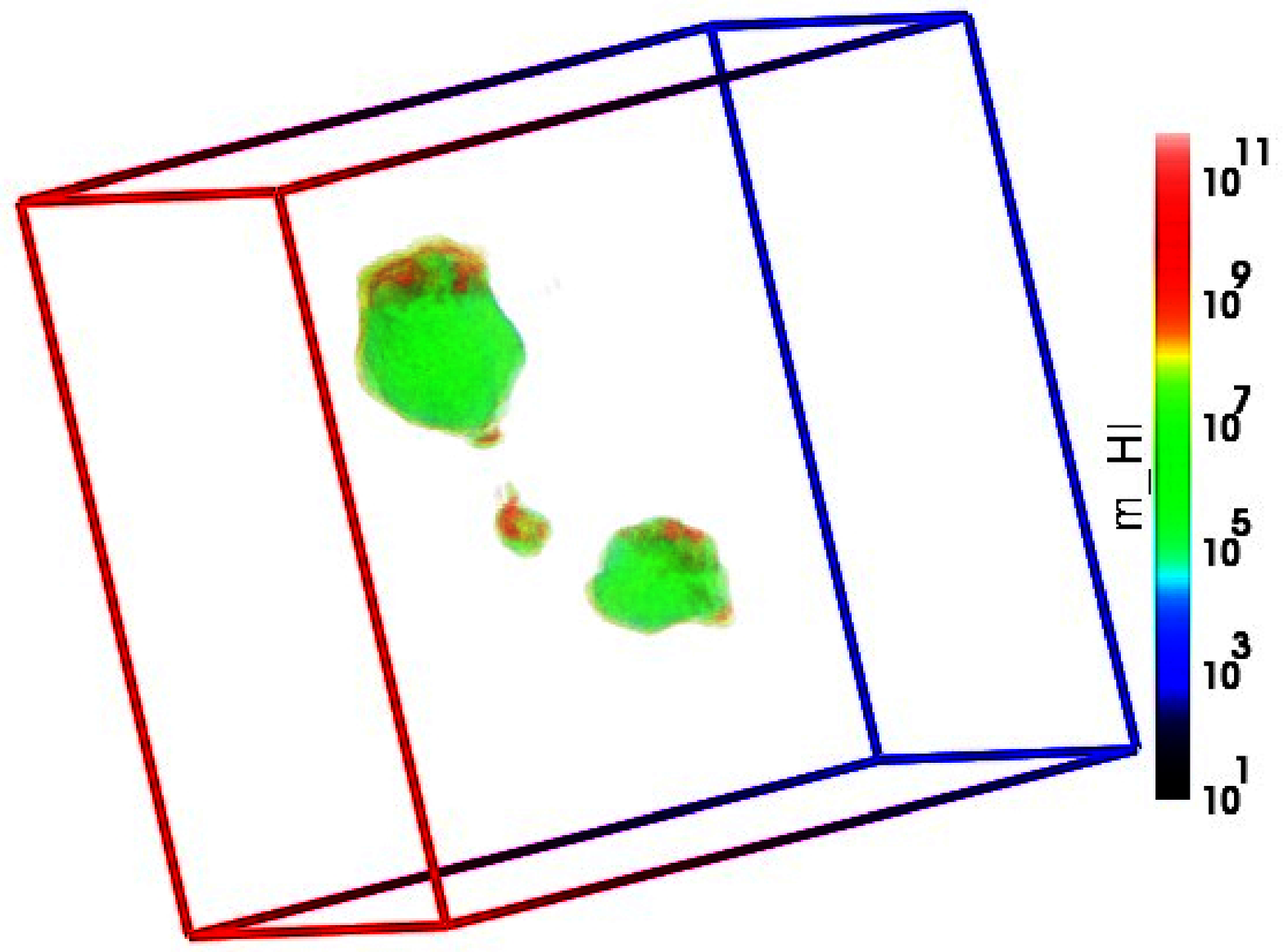} 
\includegraphics[width=2.4in]{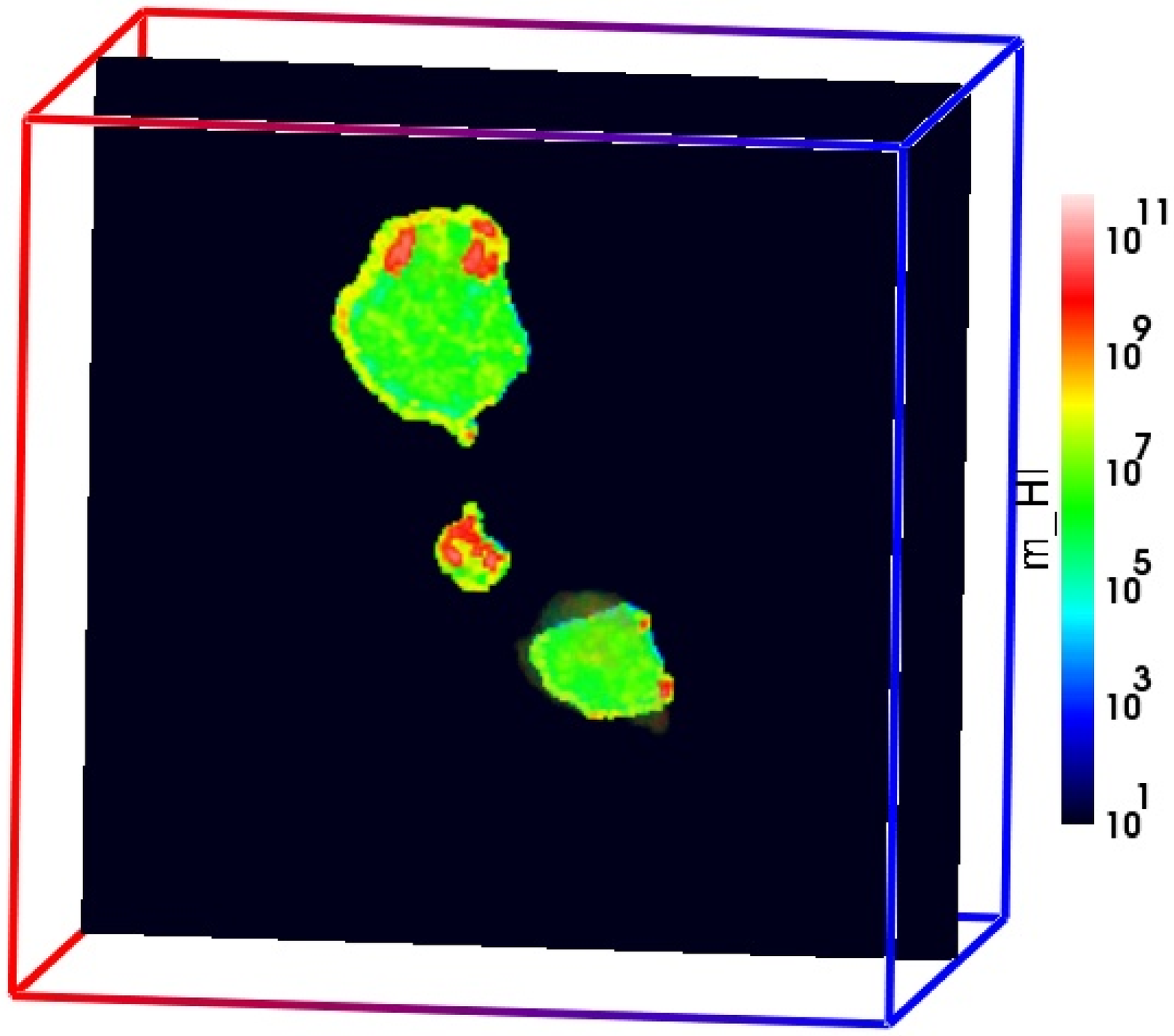} 
\includegraphics[width=2.4in]{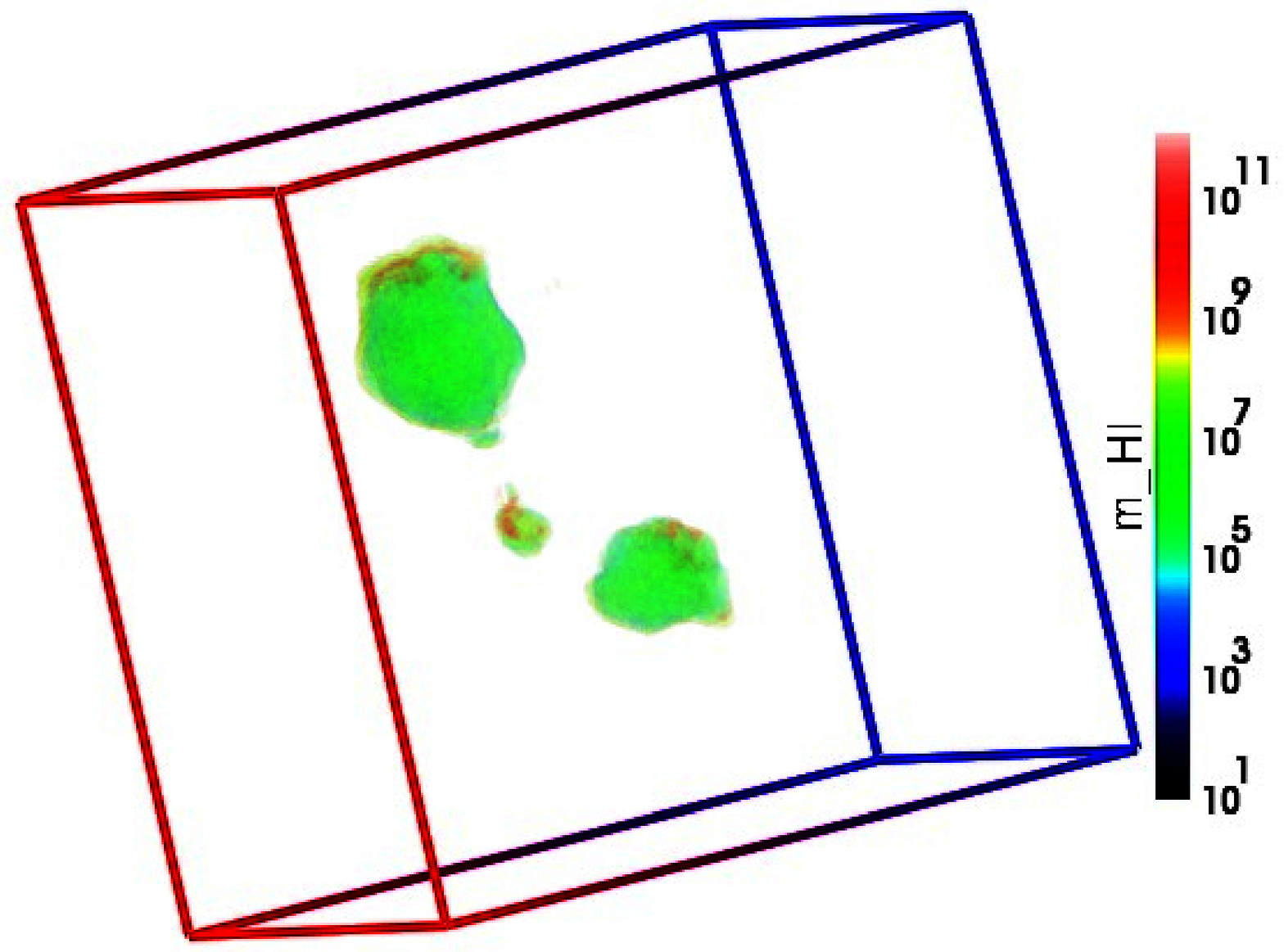} 
\includegraphics[width=2.4in]{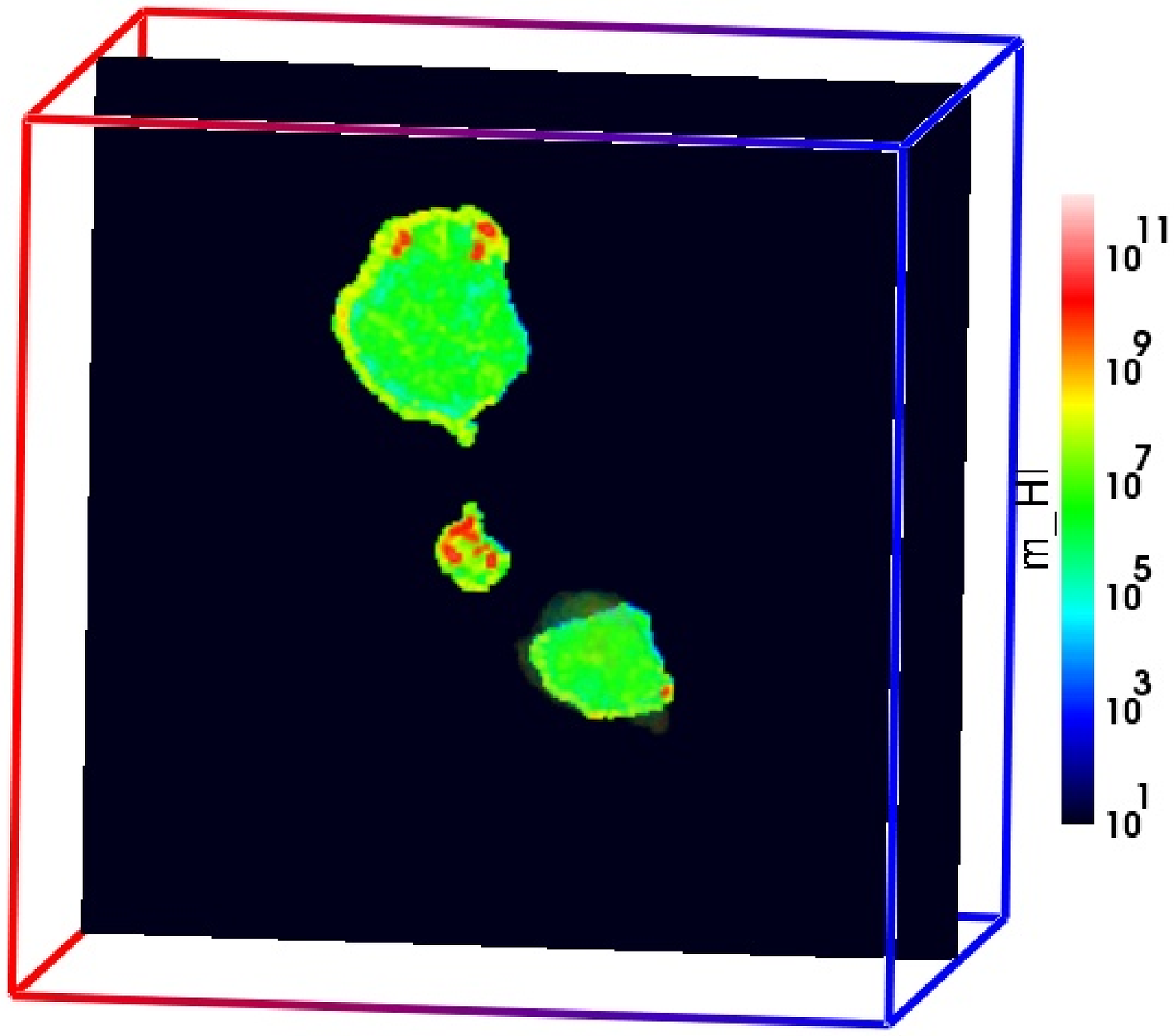} 
\includegraphics[width=2.4in]{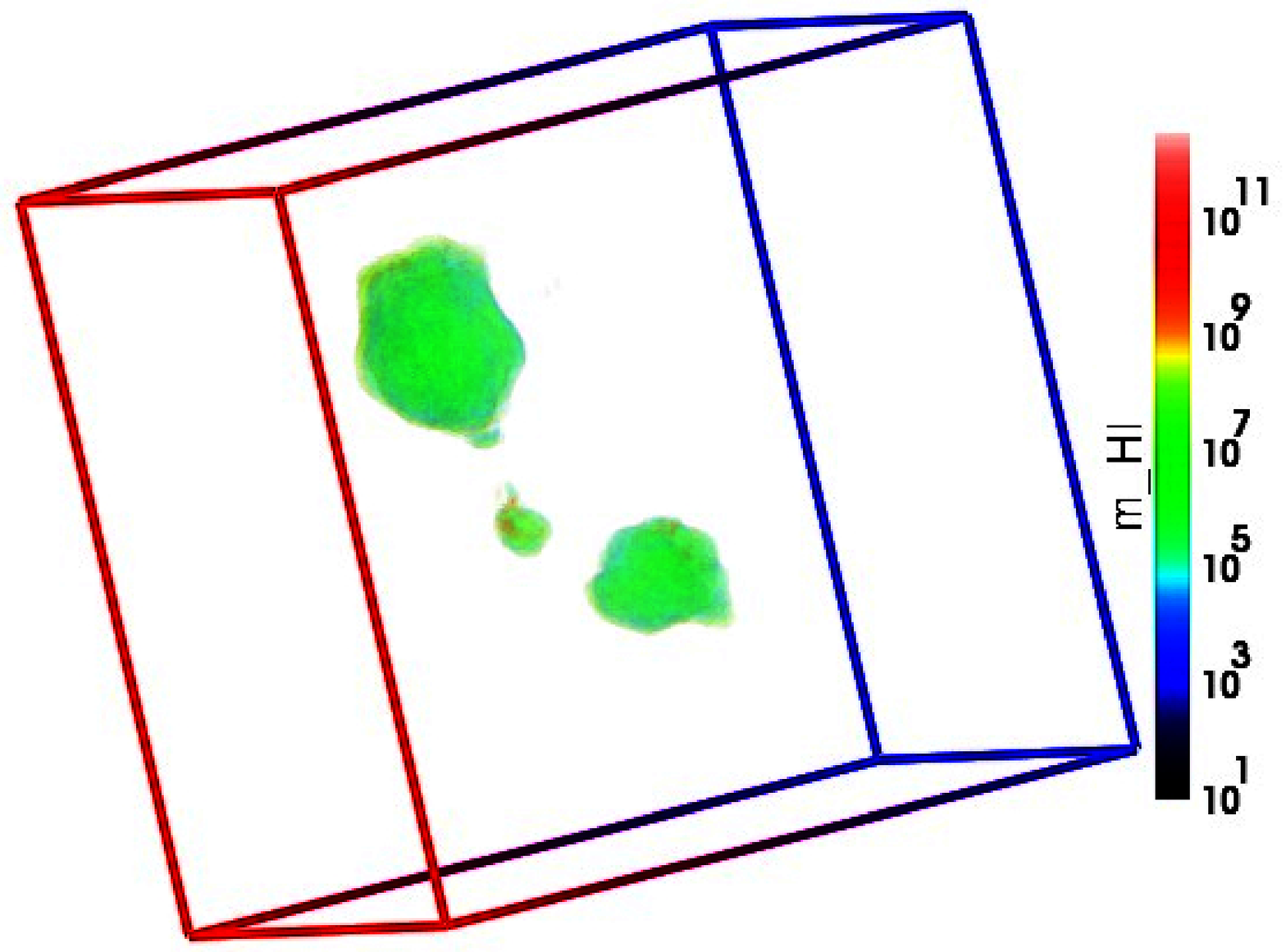} 
\includegraphics[width=2.4in]{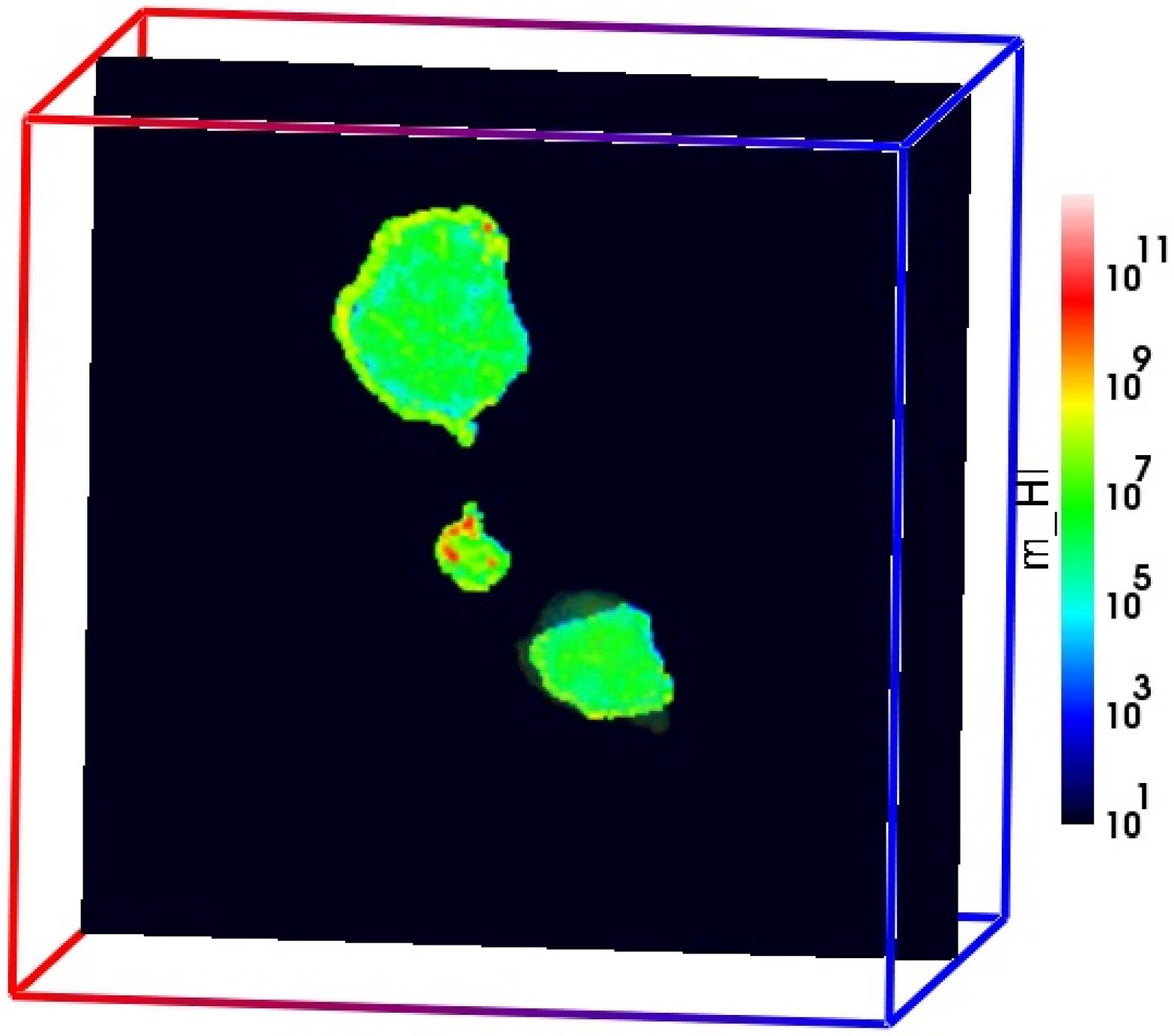} 
\includegraphics[width=2.4in]{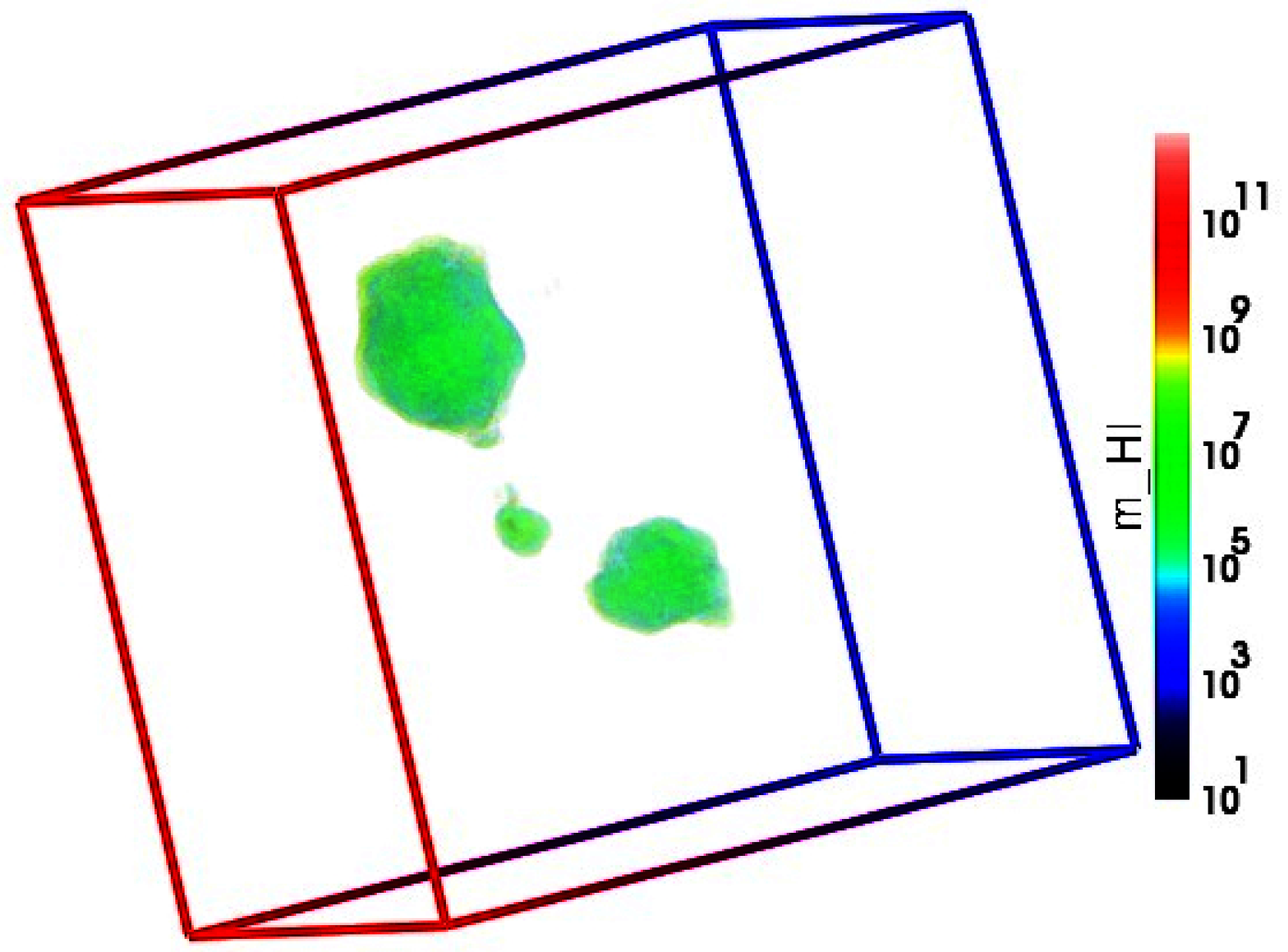} 
\includegraphics[width=2.4in]{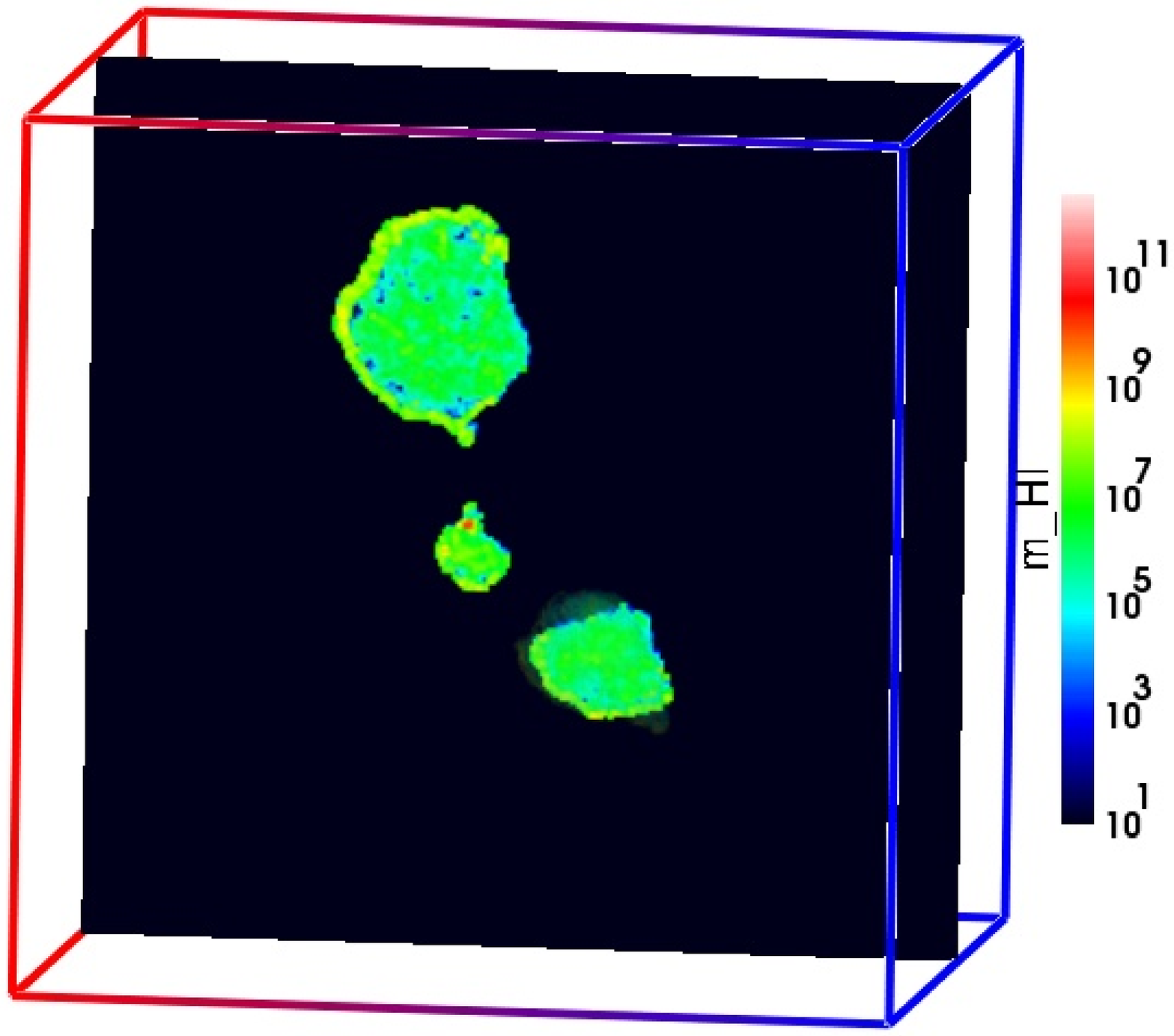} 
\caption{Evolution of the neutral mass at (top to bottom) 
redshifts $z=9,8.9,8.7,8.55$ and 8.4 in 3D volume rendering 
(left) and cross-section (right) for the Model 2 model. Red 
is neutral, green/blue is ionized.
\label{3D_cross_HI_wmap5_fig}}
\end{center}
\end{figure*}

In Figures~\ref{3D_HI_fig} and \ref{3D_cross_HI_wmap5_fig} we 
show the evolution of the neutral mass remaining in the objects 
of interest in a time sequence for simulations Model 1 and Model 
2, respectively. In each case the sequence covers the key time 
interval during which most of the LG material is ionized.
In Model 1 at $z=10.75$ the reionization of Virgo is already 
well-advanced, with a local ionized fraction above 83\%, while 
LG remains largely neutral, at less than 35\%. Shortly thereafter, 
an ionization front arrives at the proto-LG position from the 
direction of Virgo and quickly sweeps through it as illustrated 
by the next three images. This corresponds to the dramatic jump 
of the LG local ionized fraction seen in Figure~\ref{reion_hist_fig}
at $z=10.5-10.25$. By $z=10$ the Local Group material is almost 
fully ionized, reaching 97.5\% ionized fraction by mass. In this 
case, therefore the Local Group is largely ionized from the outside, 
primarily by the Virgo progenitors. 
\begin{figure*}
\begin{center}  
\includegraphics[width=3.2in]{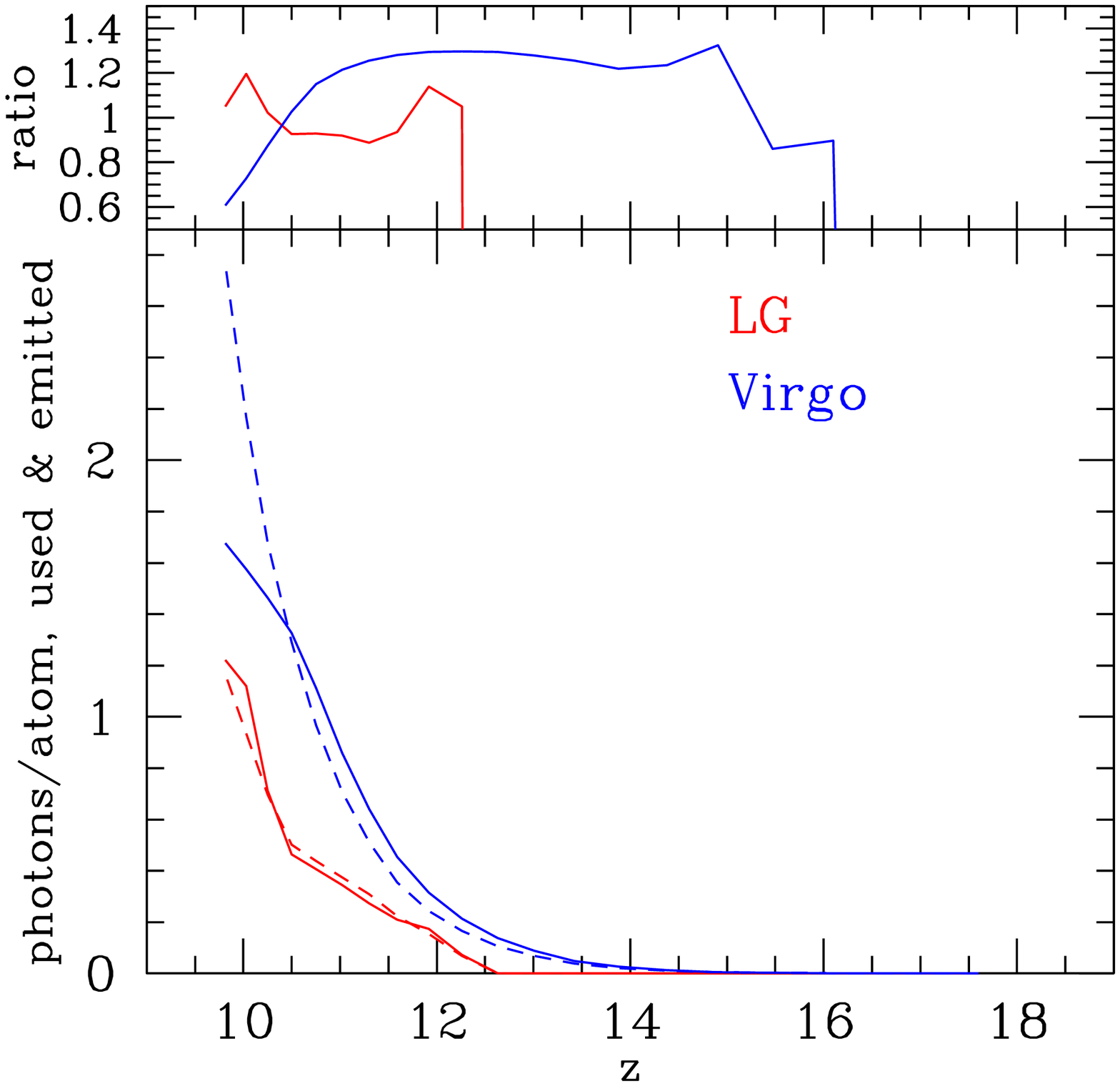} 
\includegraphics[width=3.2in]{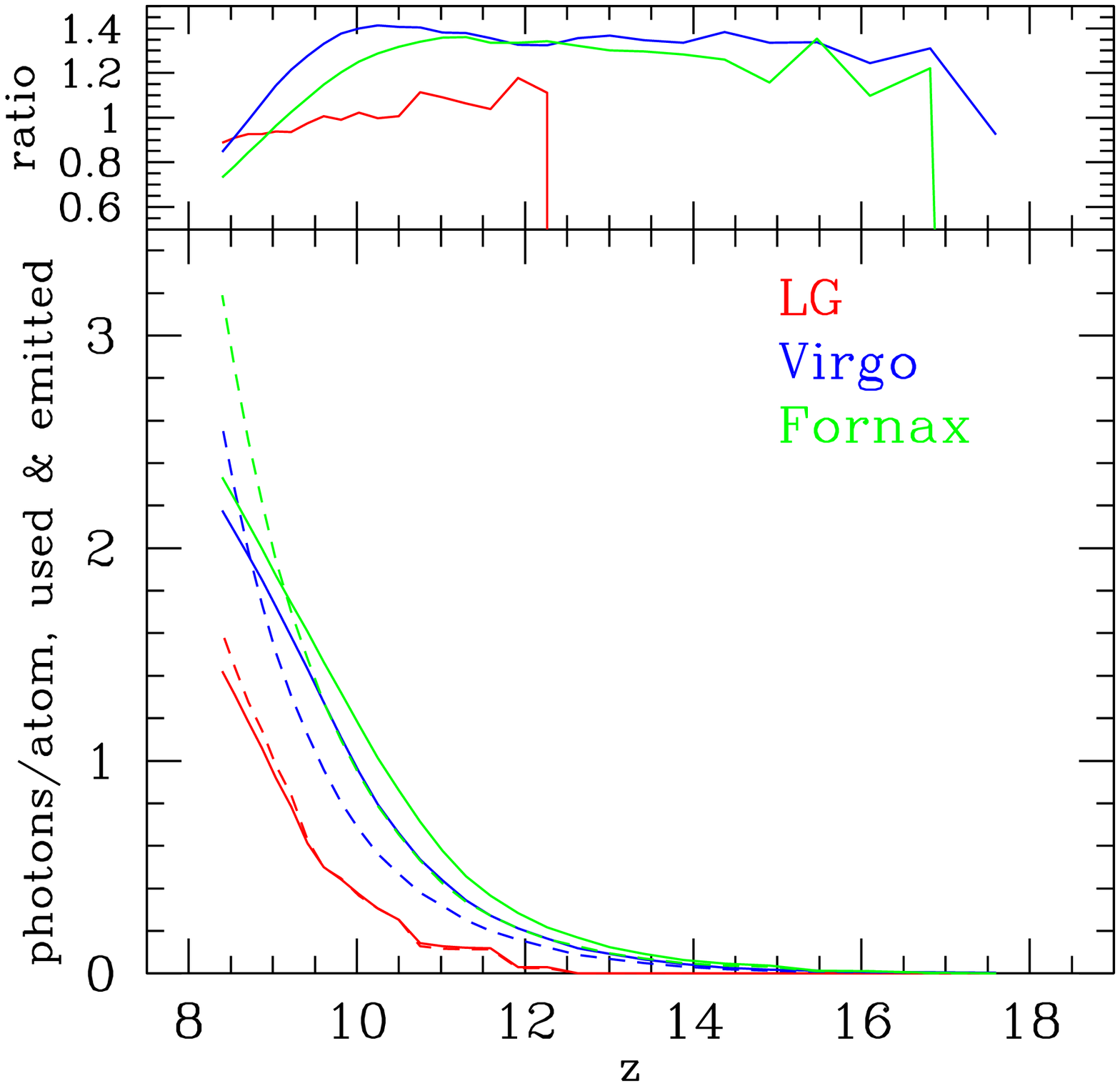} 
\caption{Cumulative number of ionizing photons for ionizations 
and recombinations (solid) and photons emitted (dashed), both 
per atom, for the Local Group and nearby clusters (as indicated 
by color) vs. redshift for Model 1 (left panel) and Model 2 
(right panel).
\label{photon_budget_fig}}
\end{center}
\end{figure*}

The evolution proceeds quite differently in the photon-poor, 
Model 2, case, as illustrated in Figure~\ref{3D_cross_HI_wmap5_fig}. 
While the nearby clusters, in this case both Virgo and Fornax, once 
again reionize themselves from the inside and relatively earlier 
than the Local Group, there are no clear ionization fronts to arrive
from them and sweep over the LG material. The Local Group internal 
sources carve ionized bubbles from the inside and eventually manage
to reionize all the LG material mostly by themselves. While we cannot
exclude modest contributions from Virgo and Fornax, the local LG 
sources appear to dominate the evolution in this case.

In order to evaulate the local reionization process for each 
structure in a more quantitative way, we counted and added up 
all ionizing photons emitted by sources within the same Lagrangean 
volume, normalized by the total number of atoms belonging to that 
object. We also counted and added together the cumulative number
of ionizing photons used up to ionize each object of interest and
also to keep it ionized (i.e. recombinations, since after each 
recombination back to neutral state that atom would need to be 
ionized again), again normalized per atom in that object. Comparing 
the values of these two numbers over time shows if that particular 
object by itself produced enough photons up to that point in time to 
fully account for its current ionization state. Results for both 
Models are shown in Figure~\ref{photon_budget_fig}.

The results confirm our conclusions based on the visual examination 
of the reionization process we discussed above. The galaxy clusters 
initially produce most of the photons needed for their own 
reionization. Until $z\sim10.5$ ($z\sim9-9.5$) for Model 1 (Model 2) 
there is some deficit of photons, i.e. a little more are used up 
than produced by the progenitors of that cluster. The reason for
this is that clusters are located at high peaks of the density 
field, and it is well established that CDM halos, and thus our 
ionizing sources, strongly cluster around such high peaks. As a 
consequence of that there are many nearby sources which surround 
the proto-cluster region and contribute to its reionization. 
However, we note that the final overlap in each case is only 
reached after the cluster itself has produced sufficient number 
of photons. We also note that recombinations have a very significant 
effect for proto-clusters, yielding usage of 1-1.5 additional 
ionizing photons per atom in addition to the one photon needed to 
achieve the initial ionization of that atom.   

On the other hand, the proto-Local Group results are different from
the cluster results in both cases. In the photon-rich Model 1 up to 
$z\sim10.5$ there are exactly as many photons produced by the LG 
progenitors as are used up for its own reionization. After that 
point however there is a significant change of slope for both curves.
Suddenly there are clearly more photons arriving than are produced 
locally. This change of slope occurs exactly when the ionization 
front from Virgo was seen above to sweep through the LG material. 
The number of photons used for ionizations and recombinations is
above the number of locally-emitted photons and it remains so up to
the redshift of full reionization (i.e. local overlap) of the Local 
group, $z\sim10$. This behaviour is more clearly seen in the top 
panel of Figure\ref{photon_budget_fig} (left), where we show the 
ratio of photons used over photons emitted locally for LG and Virgo.
This ratio starts above one at $z\sim12.3-11.7$, indicating that LG
the early LG reionization is partially helped along by nearby sources   
that do not belong to it. However, this photon ratio then falls 
slightly below one and remains so until $z=10.5$. At this point 
the external ionization front arrives and the ratio jumps to well 
above one, peaking at 1.2 and remaining above 1 until the local 
overlap. We therefore conclude that in this case most of the Local 
Group gas (up to 70\%, see Figure~\ref{reion_hist_fig}) was indeed 
reionized externally, predominantly by the nearby Virgo cluster.

In our Model 2 (photon-poor) case the photon budgets of the 
proto-clusters Virgo and Fornax are very similar to each other 
as well as to that of proto-Virgo in our Model 1. There is a 
noticeable contribution of photons from nearby clustered 
sources, which supply at least 40\% more photons than the 
respective internal sources. However, after $z\sim10$ (when
Virgo is 60\% ionized and Fornax is 70\% ionized) the ratio 
of consumed over produced photons drops quickly and by their 
full ionization ($z=8.7$) both clusters have internally 
produced more photons than are needed for their own ionization.
The local recombinations have a significant effect, resulting
in about one additional photon per atom needed to keep them
ionized.

In contast, the ionizing photons which are produced locally by 
the Local Group progenitors roughly balance the ones consumed 
for its own reionization all the way up to its neighbourhood 
overlap at $z\sim8.4$. Once again initially (at $z>11$) there 
is some external contribution of ionizing photons from nearby 
sources, but in this case throughout most of the evolution 
($8.4<z<11.5$) the emitted and consumed photons balance within 
less than 10\%. In fact, at the later stages of the evolution
the LG progenitor halos produce a bit more photons than are 
actually used to ionize it and keep it ionized. Hence, the 
Local Group evolves largely isolated in this case. After the 
local overlap the LG produces even more photons than are actually
needed to keep it ionized. The effect of recombinations on the 
LG reionization history is more modest than in the case of the 
proto-clusters, resulting in only about 0.5-0.8 additional 
photons being consumed in either simulation, in agreement with 
the lower overdensity of the LG region. Based on these results 
we therefore can conclude that in the photon-poor Model 2 the 
reionization of both the Local Group and of the nearby clusters 
is mostly a local process, with no significant external 
contributions from other structures.

\section{Summary and conclusions}

We have performed the first simulations of the reionization 
history of our local neighbourhood of the universe based on 
constrained N-body simulations of the formation of local 
structures - our Local Group of galaxies and the nearby 
galaxy clusters. The reionization history of the Local Group
cannot yet be predicted uniquely, primarily due to the still 
poor observational constraints on the properties of the 
reionization sources. Therefore, we studied two models 
constructed so as to roughly bracket the range of expected 
outcomes. While both models satisfy the available global
observational constraints, they differ significantly in
their underlying assumptions. Our first model assumes 
relatively high ionizing photon production efficiencies
('photon-rich reionization'), while our second model 
studies the other extreme, where the sources produce 
barely enough photons to complete the reionization process
in time ('photon-poor reionization'). The second model
therefore results in more extended reionization history
and delayed final overlap. The two models have the same
observational constraints imposed on their initial matter
distribution, which guarantees that the  large-scale 
structures closely resemble our local neighbourhood, but
they differ in their random component, yielding different
constrained realizations. This allows us to check the 
robustness of our results with respect to the specific 
realization. Finally, the two models also have different 
underlying cosmological models (WMAP 3-year vs. WMAP 5-year
best fit). However, the effect of the background cosmology
is well understood and results in an overall shift of the
reionization earlier or later, with no significant effects 
on our results, for which only the relative timing of 
structure formation vs. reionization history is of importance.

Our results show that the assumed efficiency of the ionizing
sources has the most important influence on the nature of the 
reionization history of our Local Group of galaxies. Efficient
photon production ensures that the nearby clusters emit more 
than sufficient number to ionize both themselves and their 
surroundings, including the Local Group. The fact that those
galaxy clusters (Virgo and Fornax) coincide with high, rare 
peaks of the density field means that they form their progenitor 
halos earlier than the LG, which is in a more average region of 
the universe. As a result, the large-scale ionization fronts 
which propagated outward from the proto-clusters overrun the
Local Group before it managed to form enough sources to ionize
itself, resulting in its reionization being mostly externally-driven.

Several points are worth noting here. Although generally the 
radiative transfer is a highly non-local phenomenon, which 
feature complicates its numerical treatment and the code 
parallelization, during most of the EoR the situation is somewhat
more complicated. The neutral patches have enormous optical depth 
to soft ionizing radiation (the only type of radiation we consider 
here). Even the already-ionized patches still have considerable 
continuum optical depth over cosmological (multiple Mpc) distances 
due to the small residual neutral fraction still remaining in such 
regions\footnote{E.g. At redshift $z=8$, mean density and residual
neutral fraction of $10^{-3}$ the mean free path is $\sim3$~Mpc. 
Density fluctuations further diminish this value.}. 
This residual neutral fraction diminishes over time, but does so 
only gradually, as more and more sources appear and the mean flux 
thereby increases. As a consequence of all this, reionization 
starts out as a fairly local process where only the relatively 
nearby, directly visible ionizing sources within the same ionized 
bubble contribute to the flux at a given point. This property allows 
us to focus our analysis on the important local sources and ignore 
the far-away ones for our current purposes (they are of course all 
included in the radiative transfer simulation). In our $\sim100$~Mpc 
box there are multiple proto-clusters which collapse nonlinearly by 
the present, but of those only Virgo and Fornax are sufficiently 
close to potentially contribute to the reionization of our Local 
Group. 

Furthermore, the ionization fronts propagate through underdense 
regions (voids) much faster than through overdense ones (filaments, 
knots). Therefore, the relative positioning of the structures of 
interest and the density fluctuations in their immediate 
neighbourhood are important. Once the available observational 
constraints are imposed in order to reproduce the local structures,
we find that the Local Group is separated from Virgo and Fornax by 
voids in either realization (see Figure~\ref{dens_slice_fig}). 
In contrast, the previous studies of this problem which did not use 
constrained realizations \citep{2007MNRAS.381..367W,2009ApJ...703L.167A}
sampled a wide range of environments and relative positions of nearby
clusters. Such, purely statistical approach yields valuable insights
on the range of reionization histories that could be expected for a
certain type object (e.g. LG-like objects). However, by its nature
such approach necessarily includes many objects which, although they
share certain basic features, locally do not reproduce the specific
large-scale structures around us. Therefore, the constrained realizations 
are indispensible if we want to make realiable predictions for the effects 
of reionization on our neighbourhood. 

Why is the mode of reionization, external vs. internal, of our Local 
Group an important issue? This has a number of important implications
for the formation of structures. Reionization dramatically rises the
Jeans mass, thus impeding the formation and growth of small galaxies. 
In terms of this effect, the galactic haloes fall into three categories.
The gas in the smallest halos (minihalos), whose virial temperatures are 
below the limit ($\sim10^4$~K) for efficient radiative cooling through 
atomic line radiation. The ionization of the gas brings its temperature 
to $\sim10^4$~K and it boils out, resulting in their complete evaporation 
\citep{2004MNRAS.348..753S,2005MNRAS...361..405I}, which leaves behind 
dark halos. In the other limit, the galaxies above certain mass 
($M\gtrsim10^{10}M_\odot$) have sufficiently deep gravitational potential 
wells to sucessfully withstand the effects of ionizing radiation and are 
thus not significantly affected by the reionization process. The effects 
of radiative feedback on dwarf galaxies of intermediate mass, roughly 
between $10^8M_\odot$ and $10^{10}M_\odot$ is more complex and still very 
much a subject of active investigation. The gas in such already-formed
systems cannot be photoevaporated, as it can cool back down to $\sim10^4$~K 
very efficiently. However, photoionization heating rises the intergalactic 
gas temperature and pressure, which rises the Jeans mass and thereby 
suppresses the future formation of very low-mass galaxies, as well as 
curtails the fresh gas infall onto such halos. Larger galaxies are less 
affected directly, but could do so indirectly, through their smaller 
progenitors, which could be expected e.g. to result in smoother gas 
sub-structure and modified stellar populations. Where the boundary 
between efficient and inefficient feedback from reionization lies is 
still unclear and very much subject of active research. Full investigation 
of the effects of reionization on galaxy formation and satellite galaxy 
populations goes well beyond the scope of the current work. However, our 
present results indicate that the photon production efficiencies of the 
first galaxies are the main factor determining the type of reionization 
history which our Local Group underwent. Therefore, this process should 
have left useful fossil records in the properties of our neighbourhood 
which will help us use local observations to answer some of the key 
questions about the young universe. 

\section*{Acknowledgments} 
This study was supported in part by Swiss National Science 
Foundation grant 200021-116696/1 and Swedish Research Council 
grant 60336701. GY acknowledges support of  MICINN (Spain) 
through research grants FPA2009-08958,  AYA2009-13875-C03-02 
and CONSOLIDER-INGENIO  SyEC (CSD2007.0050). Y.H. has been 
partially supported by the ISF (13/08). The CLUES 
simulations have been performed  in the MareNostrum supercomputer 
at BSC (Spain) and in the HLRBII Altix computer at LRZ (Germany).
We also thank DEISA for  granting us cpu time in these computers 
through two DECI projects SIMU-LU and SIMUGAL-LU. We thank Nick 
Gnedin for making publicly available his visualization code IFRIT, 
which was used to produce the images in Figs. 4-6. Some of the  
radiative transfer simulations were run on SNIC computing time at 
HPC2N (Ume\aa, Sweden). We thank Kristin Riebe for providing  Figure 2 
made with PMviewer \footnote{http://pmviewer.sourceforge.net/} 
of Arman Khalatyan.
 
\bibliographystyle{mn} 
\bibliography{paper}

\end{document}